\documentclass[12pt,twoside]{article}

\usepackage[dvips]{graphicx}
\usepackage{amssymb}

\newlength{\dinwidth}
\newlength{\dinmargin}
\newcommand{\beq}[1]{\begin{equation}\label{#1}}
\newcommand{\eeq}{\end{equation}}
\newcommand{\beqar}[1]{\begin{eqnarray}\label{#1}}
\newcommand{\eeqar}{\end{eqnarray}}
\newcommand{\Eq}[1]{Eq.~(\ref{#1})}
\newcommand{\Ab}[1]{Fig.~{\sf dummy}}
\newcommand{\Fi}[1]{Fig.~\ref{#1}}
\newcommand{\Ta}[1]{Table~\ref{#1}}
\newcommand{\Se}[1]{Section~\ref{#1}}

\newcommand{\al}{\alpha}
\newcommand{\de}{\delta}
\newcommand{\De}{\Delta}

\newcommand{\La}{\Lambda}
\newcommand{\si}{\sigma}
\newcommand{\Si}{\Sigma}
\newcommand{\ga}{\gamma}

\newcommand{\qs}{Q^2}
\newcommand{\gev}{GeV$^2$}
\newcommand{\xfnufe}{xF_3^{\nu Fe}}\newcommand{\xDbar}{x\bar{\Delta}}
\newcommand{\msbar}{\mbox{$\overline{\rm{MS}}$}\ }
\newcommand{\as}{\alpha_s}
\newcommand{\asmz}{\alpha_s(M_Z^2)}

\newcommand{\ma}[1]{\mbox{\sf #1}}
\newcommand{\csq}{\chi^2}
\newcommand{\Ftp}{F_2^p}
\newcommand{\Ftd}{F_2^d}
\newcommand{\Fdop}{F_2^d/F_2^p}
\newcommand{\Fnop}{F_2^n/F_2^p}
\newcommand{\rar}{\rightarrow}
\newcommand{\xdbmub}{x(\bar{d}-\bar{u})}
\newcommand{\Rfthr}{xF_3^{\nu Fe}/xF_3^{\nu N}}
\newcommand{\Rftwo}{F_2^{\nu Fe}/F_2^{\nu N}}

\title{
\begin{flushright}
\tt\normalsize{DESY-99-038\\
    NIKHEF-99-011} 
\end{flushright}
\vspace{1cm}
\bf A QCD analysis of HERA and fixed target structure function data
}

\author{
M. Botje\thanks{email: m.botje@nikhef.nl}\\
NIKHEF,
PO Box 41882, 1009DB Amsterdam, the Netherlands
}

\date{\today}

\begin{document}

\maketitle

\begin{abstract}

The parton momentum density distributions of the proton are determined
from a next-to-leading order QCD analysis of structure functions
measured at HERA and by fixed target experiments. Also included are data
on the difference of the up and down anti-quark distributions.  The
uncertainties in the parton densities, structure functions and related
cross sections are estimated from the experimental errors and those on
the input parameters of the fit.  Several QCD predictions based on the
parton densities obtained from this analysis are calculated and compared
to data.

\end{abstract}

\section{Introduction}\label{se:intro}

Parton momentum density distributions are important ingredients in the
calculation of high energy hadron-hadron and lepton-hadron scattering
cross sections. In these calculations the cross sections are written as
a convolution of the parton densities and the elementary cross sections
for parton-parton or lepton-parton scattering.  Whereas the latter can
be perturbatively calculated in the framework of the Standard Model, the
parton densities are non-perturbative and cannot, at present, be
calculated from theory; they are obtained from fits to a body of high
energy scattering data. In these fits measurements of deep inelastic
lepton-hadron scattering play an important r\^{o}le because in this
process the hadron structure is directly probed by the structure-less
incident lepton. 

Standard sets of parton density distributions have been published by,
among others, the groups CTEQ, MRS and GRV,
see~\cite{ref:cteq5,ref:mrstnew,ref:grv98} for their most recent
results.  These parton distribution sets are widely used as an input to
cross section calculations.  However, estimates of the uncertainties on
the parton densities are not provided. These uncertainties tend to
dominate the errors on the predicted cross sections and are an
increasingly important limitation on the interpretation of hadron-hadron
or lepton-hadron collider data. 

For instance with the integrated luminosity of about 50~pb$^{-1}$
collected at HERA during the years 1994--1997 a new kinematic domain of
large $x$ and $\qs$ becomes accessible for the study of deep inelastic
scattering in $ep$ collisions. Here $\qs = -q^2$ with $q$ the
four-momentum transfer from the incident electron to the proton and $x =
\qs/2p \cdot q$ with $p$ the four-momentum of the proton.  Measurements
by ZEUS of the $e^+p \rar e^+X$ neutral current (NC) and $e^+p \rar
\bar{\nu}X$ charged current (CC) Born cross sections have recently been
published~\cite{ref:ncnew,ref:ccnew} (see~\cite{ref:h1nccc} for recent
NC and CC cross section measurements by H1).  These data extend the
largest measured value of $\qs$ from about 5000~\gev~\cite{ref:zeusnv}
to $\qs = 5$--$6 \times 10^4$~\gev.  Standard Model predictions for the
cross sections calculated with the parton distribution set
CTEQ4~\cite{ref:cteq4} are in good agreement with the NC results but
fall below the ZEUS CC measurements at large $x$ and $Q^2$.  This could
be an indication of new physics beyond the Standard Model but might also
be due to an imperfect knowledge of the parton densities in this
kinematic region.  For instance in~\cite{ref:bodek} it is shown that a
modification of the CTEQ4 down quark density yields predictions in
agreement with the CC $e^+p$ data. 

To investigate these issues we have performed a next-to-leading order
(NLO) QCD analysis of deep inelastic structure function data from HERA
and fixed target experiments. The HERA data extend down to low values of
$x \approx 10^{-5}$ where charm mass effects must be taken into account.
For this we could have used, since it is implemented in the QCD
evolution code, the heavy quark three flavor scheme of~\cite{ref:f2c}.
However, the CC cross sections and the Z-exchange contribution to the NC
cross sections cannot be calculated in this scheme.  Because we wish to
compare the QCD predictions to the ZEUS NC and CC data at large $\qs$ we
do not use the heavy quark scheme and assume, instead, that the charmed
and bottom quarks are massless.  As a consequence, we have restricted
the kinematic domain of the analysis to larger values of $x > 10^{-3}$
where the massless approach is expected to yield reasonable QCD
extrapolations to large $\qs$, see \Se{se:compa}. 

In this paper we present the results of the QCD analysis including
estimates of the uncertainties in the parton densities, the structure
functions and the Standard Model predictions for the NC and CC cross
sections. The results of this analysis, including the full error
information, are available from the web site given in \Se{se:summary}. 

\section{QCD fit} \label{se:qcdfit}

The data used in the fit were the proton structure function ($\Ftp$)
measured at HERA by ZEUS~\cite{ref:zeusnv} and H1~\cite{ref:h1qcd}
together with the proton and deuteron ($\Ftd$) structure functions from
fixed target experiments~\cite{ref:e665,ref:newnmc,ref:bcdms,ref:slac}.
Also included in the fit were measurements of the ratio $\Fdop$ by
NMC~\cite{ref:f2dop} and neutrino data on $\xfnufe$ from
CCFR~\cite{ref:ccfrxf3}. Recent results from E866~\cite{ref:e866} on
di-muon (Drell-Yan) production cross sections in $pp$ and $pd$
scattering yields information on the difference between the down and up
anti-quark densities. Because a NLO calculation of the Drell-Yan cross
sections is not incorporated in the present analysis we have, instead,
constrained this difference by using their results on
$x(\bar{d}-\bar{u})$ as an input to the fit. 

Corrections for nuclear effects were applied to $\Ftd$, $\Fdop$ and
$\xfnufe$. For this purpose the structure function (per nucleon)
of nucleus $A$ was parameterized as
\beq{eq:kadef}
F_i^A = F_i^N [ 1 + K_A (R_A -1)]
\eeq
where $F_i^N$ is the structure function of a free nucleon (taken to be
the average of proton and neutron) as predicted by the QCD fit,
$R_A$ is a parameterization of the ratio $F_i^A/F_i^N$ taken from the
literature (see below) and $K_A$ is a parameter which controls the
size of the nuclear correction applied.

In \Fi{fig:emc}a we plot the ratio $F_2^d/F_2^N$ versus $x$ as
parameterized by Gomez et al~\cite{ref:gomez}. We allow for an error of
100\% on this nuclear correction as indicated by the shaded band in the
figure, that is, we set $K_d = 1 \pm 1$.  The nuclear correction applied
to $\xfnufe$ is shown in \Fi{fig:emc}b. For this we took the
parameterization of Eskola et al~\cite{ref:eskola} of the nuclear
correction for iron on the valence quark density $x(q-\bar{q}) \propto
xF_3^{\nu N}$.  Here we allowed for an error of 50\% i.e.\ $K_{Fe} = 1
\pm 0.5$ (shaded band in \Fi{fig:emc}b). 

The QCD predictions for the structure functions were obtained by solving
the DGLAP evolution equations~\cite{ref:dglap} in NLO in the \msbar\
scheme~\cite{ref:furm}.  These equations yield the quark and gluon
momentum densities (and thus the structure functions) at all values of
$\qs$ provided they are given as functions of $x$ at some input scale
$\qs_0$. Also required is an input value for the strong coupling
constant which, in this analysis, was set to $\asmz = 0.118$ $(\pm
0.005)$.  With a charm (bottom) threshold of $m_{c(b)} = 1.5$ (5)~GeV in
the $\qs$ evolution of $\as$ this corresponds to values of the QCD scale
parameter $\La({\msbar}) = (404,343,243)$~MeV for $f = (3,4,5)$ flavors. 

As already mentioned in the introduction, we use in this analysis the
light quark variable flavor number scheme where all quarks are assumed
to be massless and where charm and bottom are dynamically generated
above some given thresholds $\qs_c$ and $\qs_b$.  These thresholds were
set to 4 ($\pm 1$) and 30~\gev\ respectively. 

The input scale of the DGLAP evolutions was chosen to be $\qs_0 =
4$~\gev\ and the gluon distribution ($xg$), the sea quark distribution
($xS$), the difference of down and up anti-quarks ($\xDbar$) and the
valence distributions ($xu_v$, $xd_v$) were parameterized as
\beqar{eq:param}
xg(x,Q_0^2) & = & A_g x^{\de_g}(1-x)^{\eta_g} 
(1 + \gamma_g x) \nonumber \\
xS(x,Q_0^2) & \equiv &  2x ( \bar{u} + \bar{d} + \bar{s} ) =
A_s x^{\de_s}(1-x)^{\eta_s}(1 + \gamma_s x) \nonumber \\
\xDbar(x,Q_0^2) & \equiv & x(\bar{d} - \bar{u}) =
A_{\De} x^{\de_{\De}}(1-x)^{\eta_{\De}} \\
xu_v(x,Q_0^2) & \equiv & x(u-\bar{u}) = A_u x^{\de_u}(1-x)^{\eta_u} 
(1 + \gamma_u x) \nonumber \\
xd_v(x,Q_0^2) & \equiv & x(d-\bar{d}) = A_d x^{\de_d}(1-x)^{\eta_d} 
(1 + \gamma_d x). \nonumber 
\eeqar
In the fit, we assume that the strange quark distribution $x(s+\bar{s})
\equiv 2x\bar{s}$ is a given fraction $K_s = 0.20\ (\pm 0.03)$ of the
sea at the scale $\qs = 4$~\gev, in accordance with the measurements of
CCFR~\cite{ref:ccfrs}. 

The parameters $A_g$, $A_u$ and $A_d$ were fixed by the momentum sum
rule and the valence quark counting rules:
\[
\int_0^1 (xg + x\Si) \;  dx =  1,\ 
\int_0^1 u_v \; dx          =  2,\ 
\int_0^1 d_v \; dx          =  1
\]
where $x\Si \equiv xS + xu_v + xd_v$ denotes the singlet quark density
(sum of all quarks and anti-quarks).  There are thus 16 free parameters
describing the input parton densities.

From the evolved parton densities the relevant structure functions were
calculated in NLO.  In the calculations both the renormalization and the
factorization scale were set equal to $\qs$. Higher twist contributions
to $\Ftp$ and $\Ftd$ were taken into account phenomenologically by
describing these structure functions as
\beq{eq:htwist}
F_2^{\rm HT} = F_2^{\rm QCD} [ 1 + H(x)/\qs ]
\eeq

where $F_2^{\rm QCD}$ obeys the QCD evolution equations. The function
$H(x)$ was parameterized as a fourth degree polynomial in $x$ with five
free parameters:
\beq{eq:htwpar}
H(x) = h_0 + h_1 x + h_2 x^2 + h_3 x^3 + h_4 x^4.
\eeq
We assume in this analysis that $H(x)$ is the same for the proton and
the deuteron~\cite{ref:marcv} which implies that the ratio $\Fdop$ is
not affected by higher twist contributions.  We did not correct the data
for target mass effects~\cite{ref:georgi} so that these effectively are
included in~$H(x)$.  No higher twist, target mass or slow rescaling
corrections~\cite{ref:georgi,ref:barnett} were applied to $\xfnufe$. 

The normalizations of the ZEUS, H1 and NMC data sets were kept fixed to
unity whereas those of E665, BCDMS, SLAC and CCFR were allowed to float
within the quoted normalization uncertainties (7~parameters).  There are
thus in total 28 free parameters in the fit. 
 
The following cuts were made on the data:
\begin{enumerate}
\item Discard data with $x < x_{\mathrm{min}} = 0.001$. This cut was introduced
      because the massless quark scheme used in this analysis is not suitable
      to describe low $x$ structure functions where the contribution
      from charm is large. For the $xF_3$ data the cut was raised to
      $x_{\mathrm{min}} = 0.1$ to reduce the sensitivity to nuclear
      corrections at low $x$;
\item Discard data with $\qs < 3$~\gev. This cut reduces the sensitivity
      to QCD corrections beyond NLO. Because the QCD evolution of the ratio 
      $\Fdop$ is small, we apply a lower $\qs$ cut
      of 1~\gev\ to these data;
\item Discard data with $W^2 = m^2_p + \qs (1-x)/x < 7$~\gev\ ($m_p$ is the 
      mass of the proton) to reduce the 
      sensitivity to higher twist contributions and target mass effects which
      become important at high $x$ and low $\qs$.
\end{enumerate}

The NLO QCD evolutions were calculated with the program Qcdnum~\cite{ref:qcdnum};
the $\csq$ minimization and the calculation of the covariance matrices were
done with Minuit~\cite{ref:minuit}.

\section{Error propagation} \label{se:errors}

In this section we describe how the errors are defined in this analysis.
The calculation of the statistical and systematic errors is based on
linear error propagation. This implies that asymmetries in the input
errors, if any, are ignored and that the output statistical and
systematic errors are, per definition, symmetric. 

In this analysis we have incorporated the effects of the point to point
correlated experimental systematic errors in the model prediction for
the structure functions. This model prediction, calculated at the
kinematic point ($x_i,\qs_i$), is defined as
\beq{eq:sysdef}
F_i(p,s) = F_i^{\rm QCD}(p) \biggl( 1 + \sum_j  s_j \De_{ij}^{\rm syst} \biggr)
\eeq
where $F_i^{\rm QCD}(p)$ is the QCD prediction and $\De_{ij}^{\rm syst}$
is the relative systematic error on data point $i$ stemming from source
$j$.\footnote{If the systematic error is asymmetric the average of the
upper and lower error is taken.} In \Eq{eq:sysdef} $p$ denotes the set
of parameters describing the input parton densities and $s$ is a set of
systematic parameters. Notice that the experiments, by giving central
values and one standard deviation systematic errors, essentially provide
`measurements' of these systematic parameters: $s_j = 0 \pm 1$. 

Assuming that the $s_j$ are uncorrelated and Gaussian distributed with
zero mean and unit variance, the $\csq$ can be written as
\beq{eq:chi2}
\csq = \sum_{i}
 \biggl(\frac{F_i(p,s)-f_i}{ \De f_i}\biggr)^2 + \sum_{j} s_j^2
\eeq
where $f_i$ is the measured structure function and $\De f_i$ is the
statistical error with the point to point {\em uncorrelated} systematic
errors added in quadrature. 

To propagate the statistical and systematic errors two Hessian matrices
$\ma{M}$ and $\ma{C}$ were evaluated (with Minuit) at the minimum $\csq$:
\beq{eq:hessian}
M_{ij} = \frac{1}{2} \frac{\partial^2 \csq}{\partial p_i \partial p_j}
,\ \  
C_{ij} = \frac{1}{2} \frac{\partial^2 \csq}{\partial p_i \partial s_j}.
\eeq
The statistical covariance matrix of the fitted parameters is, as usual,
given by
\beq{eq:vstat}
\ma{V}^{\rm stat} = \ma{M}^{-1}.
\eeq
A systematic covariance matrix of the fitted parameters can be defined
as~\cite{ref:pascaud}
\beq{eq:vsyst}
\ma{V}^{\rm syst} =  \ma{M}^{-1} \ma{C} \ma{C}^T \ma{M}^{-1}. 
\eeq
The error on any function $F$ of the parameters $p$ (parton densities,
structure functions, cross sections etc.) is then calculated using the
standard formula for error propagation
\beq{eq:eprop}
(\De F)^2 = \sum_i \sum_j \frac{\partial F}{\partial p_i} V_{ij}
\frac{\partial F}{\partial p_j}
\eeq
where $\ma{V}$ is the statistical, the systematic, or if the total error
is to be calculated, the sum of both covariance matrices.

The sources which contribute to the error on the results of the fit are:
\begin{enumerate}
\item The statistical and systematic errors on the data. These were propagated
      as described above.
      The following systematic errors were taken into account in this
      analysis: for the ZEUS data we included the 6
      contributions as parameterized in~\cite{ref:zeusnv}. 
      The SLAC and BCDMS systematic errors
      were taken to be those used in the QCD analysis of~\cite{ref:marcv}.
      For the other data sets we included the systematic uncertainties as
      published
      (for H1 and E665 only the normalization error and the
      total systematic error are available; the CCFR systematic errors are
      given in~\cite{ref:seligman}). Apart from the 7 normalization
      constants which were left free parameters in the fit, 
      50 independent sources of systematic error were
      propagated taking into account the correlations between those of
      the NMC data sets. We assume, however, that the errors on $\Fdop$
      are independent from those on $\Ftp$ and $\Ftd$. This has
      little effect on the total error estimate since the contribution
      from $\Fdop$ is small;
\item The error on the strong coupling constant $\De\asmz=0.005$,
      on the strange quark content of the proton $\De K_s = 0.03$ and
      on the nuclear corrections to $\Ftd$, $\Fdop$ and $xF_3$
      ($\De K_d = 1$, $\De K_{Fe} = 0.5$).
      Also included is an assumed error on
      the charm threshold $\De \qs_c =1$~\gev.
      To determine these uncertainties, the
      fits were repeated with the parameter lowered or raised
      by the error and the corresponding
      error band is defined as the envelope of the results from
      the central fit and these two additional fits;
\item An additional error band (`analysis error')
      is defined as the envelope of the results from
      the central fit and several alternative fits described 
      in \Se{se:checks};
\item The renormalization scale uncertainty. Two fits were performed with
      the renormalization scale set to $\mu_R^2 = \qs/2$ and
      $\mu_R^2 = 2\qs$ while keeping the factorization scale
      $\mu^2_F$ fixed to $\qs$. The error band is defined
      as the envelope of the results from these fits and the central
      fit;
\item The factorization scale uncertainty. Here the factorization
      scale was varied in the range $\qs/2 < \mu_F^2 < 2 \qs$ while keeping 
      the renormalization scale $\mu^2_R$ fixed to $\mu^2_F$.
\end{enumerate}

Unless otherwise stated the errors shown in this paper are the quadratic
sum of the contributions~(1)--(3) listed above, that is, the scale
errors are not included. The error contributions to the parton densities
and several structure functions from each of the sources are given in
\Se{se:checks} where also the scale uncertainties are shown. 

\section{Results} \label{se:result}

The fit yielded a good description of the data. Adding the statistical
and systematic errors in quadrature a $\csq = 1537$ is obtained for 1578
data points and 28 free parameters.  The $\csq$ values for each data set
separately are listed in \Ta{tab:chisq}.  The values of the fitted
parameters are given in \Ta{tab:params}.  \Ta{tab:chisq} shows that the
$\csq = 235/147$ is rather high for the ZEUS $\Ftp$ data.  This is
caused by 7 data points, more or less randomly distributed in the
kinematic plane, which carry 84 units of $\csq$. We have verified that
removing these data points does not significantly change the results of
the analysis. 

The good agreement between the $\Ftp$ data and the QCD fit can be seen
from \Fi{fig:f2pvsx} where we show the $x$ dependence of $\Ftp$ at fixed
values of $\qs$.  In \Fi{fig:f2vsq} are plotted $\Ftp$ and $\Ftd$ from
NMC, SLAC and BCDMS versus $\qs$ for $x > 0.1$. Again, the agreement
between the data and the fit is satisfactory. The full (dotted) curves
in this plot show the QCD prediction including (excluding) higher twist
effects.  The higher twist correction $H(x)$ defined in \Eq{eq:htwist}
is plotted in \Fi{fig:htw} (full curve). This correction is negative for
$x < 0.5$ and becomes large and positive at high $x$. The same trend is
observed in the analysis of MRST~\cite{ref:mrsht} (asterisks). We refer
to \Se{se:checks} for further investigations of higher twist effects.
In the inset of \Fi{fig:htw} we compare $H(x)$ from this analysis with
the result from~\cite{ref:marcv}, averaged over proton and deuteron.  We
observe that the higher twist contributions obtained in this analysis
are substantially larger although part of the difference may be due to
the fact that the authors of~\cite{ref:marcv} have corrected the data
for kinematical higher twist contributions (target mass effects) before
extracting $H(x)$. 

The $\qs$ dependence of $\xfnufe$ for $x > 0.1$ is shown in
\Fi{fig:xf3vsq}.  There is good agreement between the data and the
prediction from the QCD fit, corrected for nuclear effects as defined in
\Se{se:qcdfit}. Also the ratio $\Fdop$ is well described as can be seen
from \Fi{fig:f2dop} where we plot the $x$ dependence of $\Fnop \equiv 2
\Fdop -1$, averaged over $\qs$~\cite{ref:f2dop}. This figure shows that
the error on the QCD prediction of $\Fnop$ rapidly increases above $x =
0.4$ indicating that the ratio of $d$ to $u$ quark densities, which is
directly related to $\Fnop$, is not well constrained by the data at
large $x$. 

We show in \Fi{fig:dmu} the density $x(\bar{d}-\bar{u})$ as a function
of $x$ at $\qs = 7.4$~\gev. There is reasonable agreement between the
E866 data and the fit. For comparison we also plot the prediction from
the CTEQ4 (CTEQ5~\cite{ref:cteq5}) parton distribution set which was
obtained before (after) the E866 data became available.  In
\Fi{fig:strange} the strange quark density from the QCD fit is compared
to the data from CCFR~\cite{ref:ccfrs} (not included in the fit).  The
good agreement supports the description of $x(s+\bar{s})$ with one fixed
parameter $K_s$ (see \Se{se:qcdfit}). The shaded bands in
\Fi{fig:strange} show that the error on the strange quark density, which
is dominated by that on $K_s$, is adequately taken into account by
assuming $\De K_s = 0.03$. 

In \Fi{fig:pdf} are plotted the parton densities from the QCD fit versus
$x$ at $\qs = 10$~\gev. Integrating the parton densities over $x$ we
find the following results for the momentum fractions carried by gluons
and quarks at $\qs = 4$~\gev: $I_g = \int_{0.001}^1 xg\; dx = 0.394 \pm
0.018$, $I_q = 0.594 \pm 0.018$ and $I_g + I_q = 0.988 \pm 0.002$
(statistical and systematic errors added in quadrature).  The error on
the last integral is much smaller than that on the first two since the
total momentum fraction was constrained to be unity in the QCD fit. This
clearly illustrates the importance of taking into account the
correlations between the errors on the parton densities. Notice that $1
- I_g - I_q = 0.012 \pm 0.002$ is the small contribution to the momentum
sum, at $\qs = 4$~\gev, from the region $x < 10^{-3}$. 

Also shown in \Fi{fig:pdf} are the parton densities from CTEQ4. There
is, within errors, good agreement. Notice, however, that $xd_v$ obtained
in this analysis is slightly shifted towards larger values of
$x$.\footnote{ A similar shift is seen in the most recent parton
distribution sets CTEQ5 and MRST~\cite{ref:mrstnew}.} Bodek and
Yang~\cite{ref:bodek} also obtain a harder $xd$ density by modifying
CTEQ4 (where the ratio $d/u \rar 0$ as $x \rar 1$) such that
\beq{eq:bodek}
d/u \rar d^{\prime}/u = d/u + B x (1+x).
\eeq
They find $B = 0.10 \pm 0.01$ which implies that $d/u \rar 0.2$ as
$x \rar 1$.

The $d/u$ ratio is shown in \Fi{fig:dou}. It is seen that the result
from the QCD fit with $B = 0$ (full curve) for $x < 0.75$ is close to
the modified CTEQ4 distribution with $B = 0.1$ (dashed curve).  If we
parameterize the $d$ quark density according to \Eq{eq:bodek} and leave
$B$ a free parameter in the fit we obtain $B = -0.02 \pm 0.01$
(statistical error), close to zero. In any case, the large error band
shown in \Fi{fig:dou} clearly indicates that the exact behavior of $d/u$
at large $x$ is presently not well constrained by the data. This ratio
might go to zero (CTEQ), to a constant (Bodek and Yang) or may even
diverge (this analysis) as $x \rar 1$. 

\section{Systematic checks} \label{se:checks}

To check the stability of the results we have performed a series of
alternative fits:

\begin{enumerate}
\item Use the quadratic sum of the statistical and systematic errors in
      the definition of the $\csq$ instead of taking statistical errors
      only;
\item Release the momentum sum rule constraint. In this fit
      the total momentum fraction
      carried by quarks and gluons is found to be 1.03, close 
      to unity. We did not perform a full error analysis for this
      fit but the uncertainty on the momentum sum is probably
      about the same as that given in \Se{se:result} 
      for the gluon and quark
      momentum fractions ($\pm 0.02$);
\item Set the value of the input scale to $\qs_0 = 7$ instead of to 4~\gev;
\item Lower the $\qs$ cut from 3 to 2~\gev\ and the $W^2$ cut
      from 7 to 5~\gev;
\item Raise the $\qs$ cut from 3 to 4~\gev\ and the $W^2$ cut
      from 7 to 10~\gev;
\item Remove the lower cut of $x_{\rm min} = 0.1$ on the $xF_3$ data.
\item Fix the normalization of the CCFR $xF_3$ data (to $N = 1.008$,
      see \Ta{tab:chisq}) and let the nuclear correction
      on $\xfnufe$ float by leaving $K_{Fe}$ a free parameter in the
      fit. This results in $K_{Fe} = 0.8$, consistent with the input value
      of $1.0 \pm 0.5$;
\item Use the parameterization of Seligman~\cite{ref:seligman}
      (see \Se{se:compa})
      instead of that from Eskola~et~al.\ to correct the $\xfnufe$ data
      for nuclear effects;
\item Leave the normalization of the NMC data free (one parameter
      for the 8 NMC data sets) so that {\em all} fixed target data are
      re-normalized with respect to HERA. This results in
      $N_{\mathrm{NMC}} = 1.007$ while the normalizations of the other
      data sets changed by less than 0.01, compared to the central fit;
\item Parameterize the $d_v$ density as given in \Eq{eq:bodek} and
      leave the parameter $B$ free in the fit. As already mentioned
      in \Se{se:result} we find 
      $B = -0.02 \pm 0.01$ (statistical error).
\end{enumerate}

All the fits listed in items (1)--(10) above yielded good descriptions
of the data with a $\csq$ per degree of freedom close to that of the
central fit. To quantify the stability of the QCD analysis we define an
`analysis error band' as the envelope of the results from the central
fit and these 10 alternative fits. 

In \Ta{tab:errors} the contributions to the relative error on several
parton densities and structure functions are given for each source
listed in \Se{se:errors}. These error contributions are calculated at
$Q^2 = 10$~\gev\ and are averaged over $x$.  In \Fi{fig:eband} we show
as a function of $x$ at $\qs = 10$ \gev\ the contributions to the
relative error $\De g/g$ (left-hand plots) and $\De \Si / \Si$
(right-hand plots). It is seen that the experimental errors shown in
\Fi{fig:eband}a are roughly equal to the input errors which, in
\Fi{fig:eband}b, are calculated as the quadratic sum of the
contributions listed in item (2) of \Se{se:errors}. Compared to these
two contributions the analysis error band is small (\Fi{fig:eband}c).
This illustrates the stability of the fit with respect to the ten
alternatives listed above. The factorization and renormalization scale
uncertainties added in quadrature are shown in \Fi{fig:eband}d. For the
gluon distribution the scale error is roughly equal to the other error
contributions. For the quarks however the scale error, which is
dominated by the factorization scale uncertainty, is the largest
contribution to the total error. 

To check the higher twist contributions obtained from the central fit,
we have determined these in an alternative way as follows.  First, we
performed a fit with the $\qs$ and $W^2$ cuts raised to 10 and 35~\gev\
respectively. This restricts the kinematic region to a domain where
higher twist effects are small ($<4$\% as estimated from the results of
the central fit) so that the $\qs$ evolution of the data can be
described by perturbative QCD only. Keeping $H(x)$ to zero in the fit, a
good description of the data was obtained with a $\csq = 824$ for 901
data points and 21 free parameters.\footnote{In addition to the five
higher twist parameters the normalizations of the SLAC data sets were
kept fixed because these data were almost entirely removed by the larger
cuts.} The parton densities obtained from this fit are, within errors,
consistent with the central result. 

Second, we used the parton densities obtained above to calculate the QCD
predictions for $\Ftp$ and $\Ftd$ down to the standard cuts of $\qs = 3$
and $W^2 = 7$~\gev. Keeping $F_2^{\rm QCD}$ in \Eq{eq:htwist} fixed to
these predictions we determined $H(x)$ from a fit to the data on $\Ftp$
and $\Ftd$. The result is compatible with that obtained from the central
fit as shown by the dashed curves in \Fi{fig:htw}. We conclude from this
investigation that no significant bias is introduced by including, in
the central fit, higher twist effects as parameterized in
\Eq{eq:htwist}. 

Finally, we performed a fit (with the standard cuts on the data) where
the momentum sum rule constraint was removed and where in addition the
higher twist contribution $H(x)$ was kept fixed to zero.  This resulted
in a considerably worse $\csq = 1918$ for 1578 data points and 24 free
parameters. Furthermore the total momentum fraction carried by quarks
and gluons was found to be 1.20.  This is so because a large gluon
density is favored by the large slopes of the $F_2$ data at intermediate
$x$ and low $\qs$ (see \Fi{fig:f2vsq}). We reject this fit as an
alternative to the central result because of the large $\csq$ and the
violation of the momentum sum rule. 

\section{Comparisons} \label{se:compa}

In this section we calculate a few predictions based on the parton
densities obtained from the QCD analysis and compare these to deep
inelastic scattering data which are not included in the fit (except
$\xfnufe$ at large $x$). 

First, we investigate nuclear effects on the $\nu$-Fe structure
functions by calculating the ratio of the CCFR data to the predictions
from the QCD fit of neutrino scattering on a free nucleon.  We find
that the ratios $\Rfthr$ and $\Rftwo$ do not, within errors, depend on
$\qs$ in accordance with measurements of nuclear effects in charged
lepton scattering where also no significant $\qs$ dependence has been
observed~\cite{ref:nmcnuc}. We have therefore fitted the $xF_3$ data to
the parameterization
\[
\xfnufe(x_i,\qs_j) = R_i \; xF_3^{\nu N}(x_i,\qs_j)
\]
where $xF_3^{\nu N}$ was kept fixed to the QCD prediction and $R_i
\equiv \Rfthr(x_i)$ was left a free parameter for each bin of $x$. In the
fit only statistical errors on the data were taken into account. The
upper (lower) systematic error on $R_i$ was estimated by offsetting the
data with each source of systematic error in turn, repeating the fit and
adding the deviations above (below) the central value in quadrature.
Likewise the ratio $\Rftwo$ was calculated for each bin of $x$. In the
fits only data were included above the charm threshold $\qs_c = 4$~\gev. 
 
In \Fi{fig:emcrat} the results are plotted as functions of $x$ for
$\xfnufe/xF_3^{\nu N}$ (top) and $F_2^{\nu Fe}/F_2^{\nu N}$ (bottom).
Both these ratios exhibit the typical $x$ dependence of nuclear effects
including a hint for the rise at large $x$ due to Fermi motion in the
nucleus.  It is clear that nuclear effects in neutrino scattering on
heavy targets are present and that corrections must be applied if these
data are included in QCD fits. Also shown in \Fi{fig:emcrat} are two
parameterizations of nuclear corrections from Eskola et al.~(full
curves)\footnote {Plotted are the nuclear corrections for iron on
$xu_v+xd_v \propto xF_3^{\nu N}$ (see \Se{se:qcdfit}) and $x\Si \propto
F_2^{\nu N}$.} and from Seligman~\cite{ref:seligman} (dashed curves).
Both these parameterizations are obtained from fits to data on nuclear
effects in charged lepton scattering.  To show the contribution to the
errors on $\Rfthr$ and $\Rftwo$ from the QCD fit we have chosen to draw
these as the shaded bands around the full curves.  We conclude from
\Fi{fig:emcrat} that, in the kinematic range explored and within the
present errors, there is no strong indication for nuclear effects being
different in neutrino or charged lepton deep inelastic scattering. 

Next, we investigate to which extent the massless quark scheme used in
this analysis is able to describe the charm contribution $F_2^c$ to the
inclusive $F_2$ structure function. In \Fi{fig:fcvsq} we show the recent
data from ZEUS on $F_2^c$~\cite{ref:zeusfc} versus $\qs$ for $x \geq 5
\times 10^{-4}$.  The full curves with the shaded error bands
correspond to the prediction of $F_2^c$ from the QCD fit. Because these
predictions are calculated with the assumption that the charmed quark is
massless, the curves extrapolate to $F_2^c = 0$ at the charm threshold
$\qs_c = 4$~\gev\ in disagreement with the data.  For comparison we also
plot in \Fi{fig:fcvsq} the result from the ZEUS QCD analysis
of~\cite{ref:zeusqcd} where charm mass effects were taken into account
using the heavy quark three flavor scheme as defined in~\cite{ref:f2c}
(dashed curves). This heavy quark scheme gives a good description of
$F_2^c$ over the full $\qs$ range down to the lowest measured value of
$x = 5 \times 10^{-5}$, see~\cite{ref:zeusfc}.  However, it is seen from
\Fi{fig:fcvsq} that above $\qs \approx 10$~\gev\ also the massless quark
scheme describes the data well. This supports the assumption made in the
remainder of this section that, at least for $x > 10^{-3}$, quark mass
effects do not spoil the QCD extrapolations to large $\qs$. 

To compare with the recent $e^+p$ data from
ZEUS~\cite{ref:ncnew,ref:ccnew} we have used the parton densities from
this analysis to calculate the Born cross sections for the processes
$e^+p \rar e^+X$ (NC) and $e^+p \rar \bar{\nu}X$ (CC).  For unpolarized
positrons and protons these cross sections are, at HERA energies, to
good approximation given by
\beqar{eq:xsecs}
\frac{d^2 \si^{NC} (e^+p)}{dxd\qs} & = & \frac{2\pi \al^2}{xQ^4}
\left[ Y_+ F_2^{NC} - y^2 F_L^{NC} - Y_- xF_3^{NC}
\right] \\
\frac{d^2 \si^{CC} (e^+p)}{dxd\qs} & = & \frac{G^2_F}{4 \pi x}
\ \frac{M_W^4}{(\qs + M^2_W)^2}
\left[ Y_+ F_2^{CC} - y^2 F_L^{CC} - Y_- xF_3^{CC}
\right]. \nonumber
\eeqar
Here $\al$ is the fine structure constant, $G_F$ the Fermi constant and
$M_W$ the mass of the $W$ boson. In \Eq{eq:xsecs}, $y = \qs/xs$ and
$Y_{\pm} = 1 \pm (1-y)^2$ with $s$ the square of the $e^+p$ center of
mass energy. The structure functions $F_2$, $F_L$ and $xF_3$ were
calculated as functions of $x$ and $\qs$ in NLO using the expression
given in~\cite{ref:martinelli} for $F_L$. In the NC case contributions
from $Z^0$ exchange and $\ga$-$Z^0$ interference were taken into
account. We refer to~\cite{ref:ncnew} and~\cite{ref:ccnew} and
references therein for further details concerning the calculation of the
NC and CC structure functions and for the input values of the
electroweak parameters. We mention here that in leading order QCD the CC
cross section can be written as~\cite{ref:ukatz}
\beq{eq:cclo}
\frac{d^2 \si^{CC}(e^+p)}{dx d\qs} = \frac{G^2_F}{2 \pi x}
\frac{M^4_W}{(\qs+M_W^2)^2}
\left[ (1-y)^2(xd+xs)+x\bar{u}+x\bar{c} \right]
\eeq
from which it is seen that this cross section is predominantly
determined by the down quark density. 

In \Fi{fig:sgnc} we plot the ZEUS data on the NC $e^+p$ single
differential cross sections $d \si/d\qs$, $d \si/dx$ and $d \si/dy$ for
$\qs > 400$~\gev\ together with the predictions from the QCD fit (full
curves).  Here both the data and the predictions are divided by the
cross sections calculated from the CTEQ4D parton distribution set. This
figure shows that the present analysis and CTEQ4 give a good description
of the measurements although there is a slight tendency ($\sim 5$\%) of
the QCD predictions to be below the data at low $\qs$, low $x$ and large
$y$.  These are the regions where the calculation of the cross sections
might be sensitive to higher order QCD corrections on
$F_L$~\cite{ref:zijlstra}. 

The ZEUS results on the CC $e^+p$ single differential cross sections for
$\qs > 200$~\gev\ are shown in \Fi{fig:sgcc} (again divided by the CTEQ4
predictions). The agreement between the data and the predictions from
the QCD fit (full curves) is excellent. The larger cross section at
large $x$ and $\qs$, compared to CTEQ4, is due to the harder $d_v$
distribution obtained in this analysis, see \Fi{fig:pdf}.  We have
verified that using the more recent parton distribution sets CTEQ5 and
MRST, also with a harder $d$ density, yields results close to those
presented here. 

\begin{figure} [p!]
\begin{center}
\includegraphics[bb = 50 140 530 700, scale = 0.9]{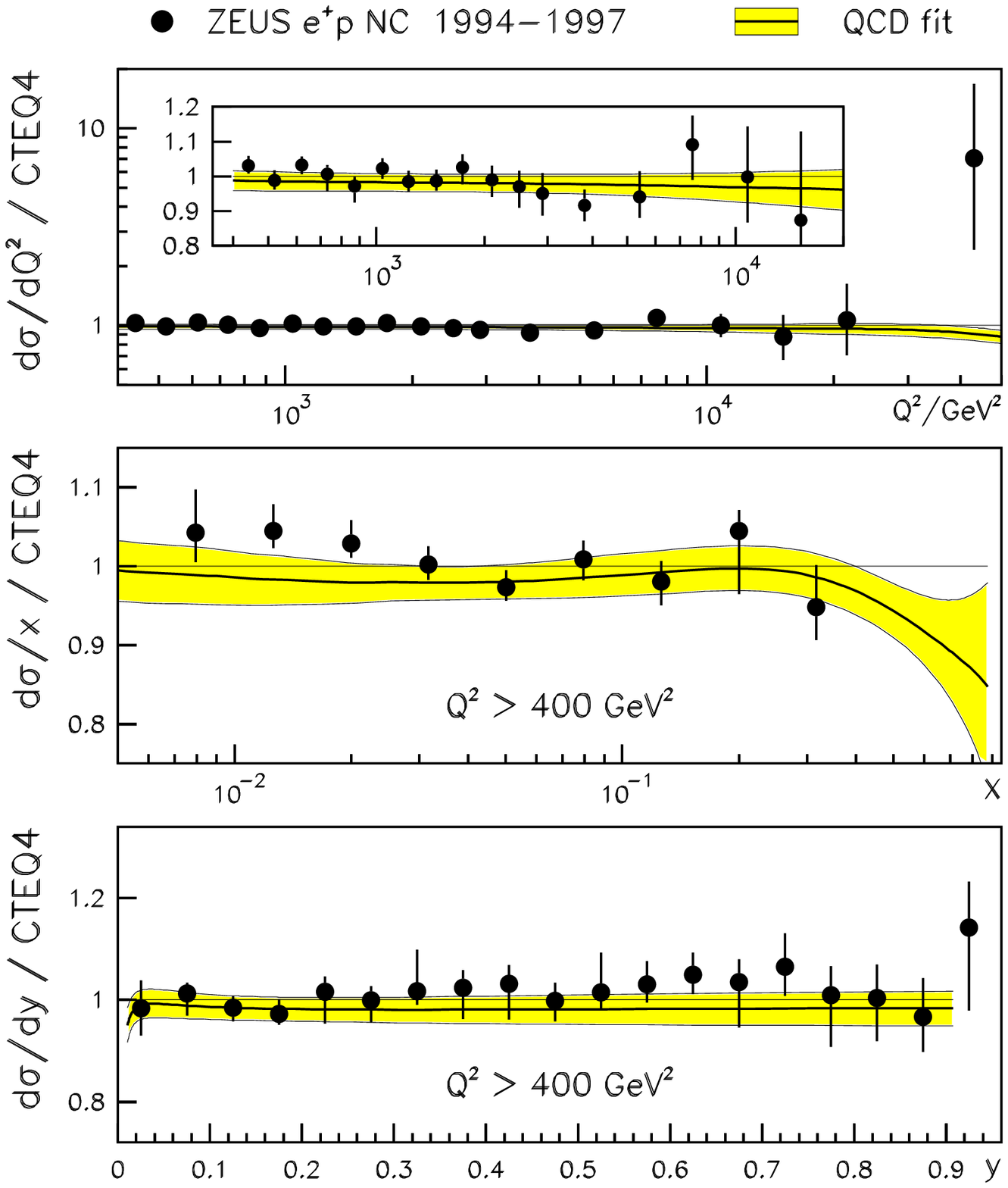}
\end{center}
\caption{
The $e^+p$ NC single differential cross sections $d\si /d\qs$ (top), $d\si
/dx$ (middle) and $d\si /dy$ (bottom) measured by ZEUS~\cite{ref:ncnew},
divided by the QCD prediction calculated with the
CTEQ4D~\cite{ref:cteq4} parton distribution set. The full curves with
shaded error bands correspond to the results of the QCD fit. The inset
in the top plot shows $d\si/d\qs$ for $\qs < 2 \times 10^4$~\gev. 
}
\protect\label{fig:sgnc}
\end{figure}

\begin{figure} [p!]
\begin{center}
\includegraphics[bb = 50 140 530 700, scale = 0.9]{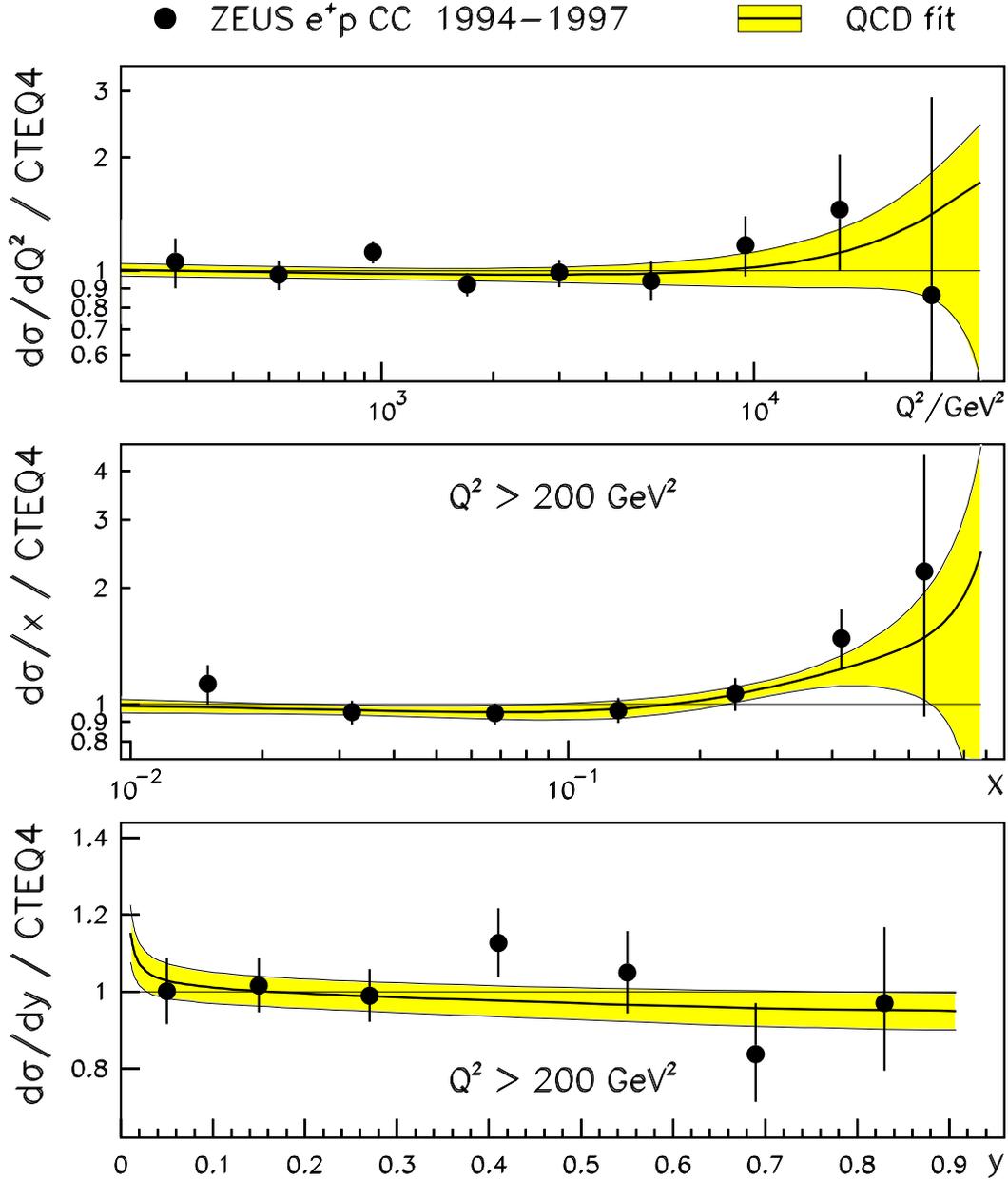}
\end{center}
\caption{
The $e^+p$ CC single differential cross sections $d\si /d\qs$ (top), $d\si
/dx$ (middle) and $d\si /dy$ (bottom) measured by ZEUS~\cite{ref:ccnew},
divided by the QCD prediction calculated with the
CTEQ4D~\cite{ref:cteq4} parton distribution set. The full curves with
shaded error bands correspond to the results of the QCD fit.
}
\protect\label{fig:sgcc}
\end{figure}

We conclude from Figs.~\ref{fig:sgnc} and \ref{fig:sgcc} that, within
the present accuracy, no significant deviations can be observed between
the data and the Standard Model predictions for $\qs$ up to about $3
\times 10^4$~\gev. 

\section{Summary} \label{se:summary}

The parton momentum density distributions in the proton were determined
from a NLO QCD analysis of structure functions measured at HERA and by
fixed target experiments, together with measurements of the difference
$x(\bar{d}-\bar{u})$.  The data included in the fit cover a kinematic
range of $0.001 < x < 0.75$ , $3 < \qs < 5000$~\gev\ and $W^2 > 7$~\gev.
The structure functions measured in nuclear targets were corrected for
nuclear effects.  Higher twist contributions to the proton and deuteron
$F_2$ structure functions were taken into account phenomenologically.
The fit was performed in the variable flavor number scheme where the
charm and bottom quarks are assumed to be massless.  It turns out that
this scheme gives for $x > 10^{-3}$ and $\qs > 10$~\gev\ a good
description of the charm structure functions measured at HERA. 

The uncertainties in the parton densities, structure functions and
related cross sections were estimated from the experimental statistical
errors, from the 57 independent sources of systematic uncertainty and
from the errors on the input parameters of the fit. An additional
analysis error is defined as the envelope of the results from the
central fit and 10 alternative fits. This error gives a small
contribution to the total error.  Also estimated are the renormalization
and the factorization scale uncertainties. 

The fit yields a good description of the data. The higher twist
contributions to $\Ftp$ and $\Ftd$ are found to be significant over the
whole $x$ range covered in this analysis. The ratio of down to up quark
densities at large $x > 0.4$ is ill constrained by the data. As a
consequence we find little sensitivity to a recently proposed
parameterization of the $d$ quark density aimed at modifying $d/u$ at
high $x$. 

Nuclear effects in neutrino scattering on iron were investigated by
comparing measurements of $\xfnufe$ and $F_2^{\nu Fe}$ with the QCD
predictions of neutrino scattering on a free nucleon. We do not, within
the present errors, find a significant difference between nuclear
effects in neutrino and charged lepton scattering. 

From the evolved parton densities the $e^+ p$ NC and CC structure
functions and Born cross sections were calculated.  There is good
agreement between these Standard Model predictions and the measurements
by ZEUS of the $e^+ p$ NC and CC cross sections at large $x$ and $\qs$. 

The results of this analysis are stored on a computer readable file
which is available from the web site {\tt
http://www.nikhef.nl/user/h24/qcdnum}.  This file contains the
statistical and systematic covariance matrices and, as functions of $x$
and $\qs$, the parton densities as well as the derivatives of these
densities to the fitted parameters. Also stored are the results of the
systematic fits given in \Se{se:errors} item (2)--(5) as well as the
analysis error band defined in \Se{se:checks}.  A Fortran
program~\cite{ref:epdflib} is provided to read the file and it contains
tools to propagate the errors to any function of the parton densities. 

\setcounter{secnumdepth}{0}
\section{Acknowledgments}

I would like to thank M.~Cooper-Sarkar, R.~Devenish, P.~Kooijman,
M.~Kuze, E.~Laenen and G.~Wolf for discussions and comments on the
manuscript. I am grateful to A.~Vogt for many detailed comparisons of
the GRV and Qcdnum evolution codes.

%-----------------------------------
\begin{table}[p!]
\begin{center}
\begin{tabular*}{15cm}{ll@{\extracolsep{\fill}}rrcl}
Dataset & $F$ & $\csq$ & points & $\csq/$point & norm.\\
\hline
 SLAC P    & $\Ftp$    &   61.3    &   56     &  1.10 & 0.988    \\
 SLAC D    & $\Ftd$    &   52.3    &   57     &  0.90 & 0.984    \\
 BCDMS P   & $\Ftp$    &  155.2    &  177     &  0.88 & 0.979    \\
 BCDMS D   & $\Ftd$    &  159.3    &  159     &  1.00 & 1.000    \\
 NMC E090 P& $\Ftp$    &   46.0    &   44     &  1.04 &          \\
 NMC E120 P& $\Ftp$    &   63.2    &   53     &  1.19 &          \\
 NMC E200 P& $\Ftp$    &   75.4    &   64     &  1.18 &          \\
 NMC E280 P& $\Ftp$    &   65.2    &   72     &  0.91 &          \\
 NMC E090 D& $\Ftd$    &   47.2    &   44     &  1.07 &          \\
 NMC E120 D& $\Ftd$    &   43.5    &   53     &  0.82 &          \\
 NMC E200 D& $\Ftd$    &   47.6    &   64     &  0.74 &          \\
 NMC E280 D& $\Ftd$    &   51.4    &   72     &  0.71 &          \\
 E665 P    & $\Ftp$    &   53.4    &   41     &  1.30 & 1.017    \\
 E665 D    & $\Ftd$    &   44.6    &   41     &  1.09 & 1.001    \\
 ZEUS      & $\Ftp$    &  235.1    &  147     &  1.60 &          \\
 H1        & $\Ftp$    &   98.2    &  150     &  0.65 &          \\
 NMC       & $\Fdop$   &  189.1    &  205     &  0.92 &          \\
 CCFR      & $\xfnufe$ &   33.2    &   68     &  0.49 & 1.008    \\
 E866      & $\xdbmub$ &   16.9    &   11     &  1.54 &          \\
\hline
 Total     &           & 1537.2    & 1578     &  0.97 &          \\
\end{tabular*}
\end{center}
\caption{
The contribution to the total $\csq$ from each data set included in the
QCD fit. The $\csq$ values are calculated with the statistical and
systematic errors added in quadrature. The last column gives the values
of the normalization constants left free in the fit. 
}
\label{tab:chisq}
\end{table}
%-----------------------------------

%-----------------------------------
\begin{table}[p!]
\begin{center}
\begin{tabular}{c r@{$\,\pm\,$}l r@{$\,\pm\,$}l
 r@{$\,\pm\,$}l r@{$\,\pm\,$}l r@{$\,\pm\,$}l       }
Parameter & \multicolumn{2}{c}{$xg$} & \multicolumn{2}{c}{$xS$} &
\multicolumn{2}{c}{$\xDbar$}&\multicolumn{2}{c}{$xu_v$}&
\multicolumn{2}{c}{$xd_v$}\\
\hline
 $A$     &\multicolumn{2}{l}{\hspace{3.2mm}1.32}&   0.81&0.03& 0.31&0.10
         &\multicolumn{2}{l}{\hspace{3.2mm}2.72}
         &\multicolumn{2}{l}{\hspace{3.2mm}1.98} \\
 $\de $  &$-$0.26&0.03&$-$0.15&0.01& 0.57&0.09
         &   0.62&0.02&   0.65&0.02    \\
 $\eta$  &   5.19&1.53&   3.96&0.19& 7.47&1.00
         &   3.89&0.02&   3.07&0.18    \\
 $\gamma$&$-$0.52&1.42&$-$1.32&0.06&\multicolumn{2}{c}{}
         &   2.79&0.31&$-$0.82&0.12    \\
         \multicolumn{11}{c}{}                 \\
&\multicolumn{2}{c}{$h_0$}&\multicolumn{2}{c}{$h_1$}&\multicolumn{2}{c}{$h_2$}
&\multicolumn{2}{c}{$h_3$}&\multicolumn{2}{c}{$h_4$}\\
\cline{2-11}
          & $-$0.12&0.06 & $-$1.9&1.1& 12.7&5.7
          & $-$45.0&12.7& 55.2&9.9\\
\end{tabular}
\end{center}
\caption{The values of the parameters obtained from the QCD fit. The
parameters are defined in \Se{se:qcdfit}. The errors given are
statistical only. The full statistical and systematic error matrices can
be accessed as described in \Se{se:summary}.}
\label{tab:params}
\end{table}

%-----------------------------------
\begin{table}[p!]
\begin{center}

\begin{tabular*}{15cm}{l@{\extracolsep{\fill}}rrrrrrrrrrr}
$f$ & stat. & syst. & $\as$ & $K_s$ & $K_d$ & $K_{Fe}$ &
      $\qs_c$ & anal. & total & $\mu^2_R$ & $\mu^2_F$ \\
\hline
          &    &     &     &     &     &     &     &     &     &     &     \\
$xg$      &3.1 & 7.1 &12.1 & 0.1 & 1.3 & 1.8 & 3.3 & 4.7 &16.1 &10.1 & 8.5 \\
$x\Si$    &0.7 & 3.9 & 1.2 & 1.2 & 0.3 & 0.8 & 2.5 & 1.6 & 5.5 & 0.6 & 5.8 \\
$xS$      &1.1 & 5.4 & 1.4 & 1.6 & 0.6 & 1.7 & 3.5 & 2.5 & 8.1 & 0.9 & 7.9 \\
$xu_v$    &0.9 & 3.4 & 1.5 & 0.0 & 1.3 & 3.6 & 0.1 & 1.0 & 5.7 & 2.1 & 1.4 \\
$xd_v$    &1.7 & 5.8 & 1.9 & 0.0 & 2.3 & 6.4 & 0.3 & 4.2 &11.0 & 2.6 & 1.4 \\
          &    &     &     &     &     &     &     &     &     &     &     \\
$\Ftp$    &0.3 & 3.3 & 0.1 & 0.0 & 0.1 & 0.2 & 0.1 & 0.8 & 3.4 & 0.2 & 0.3 \\
$\Ftd$    &0.3 & 3.3 & 0.1 & 0.0 & 0.4 & 0.3 & 0.2 & 0.8 & 3.5 & 0.2 & 0.3 \\
$\Fdop$   &0.1 & 0.4 & 0.0 & 0.0 & 0.3 & 0.1 & 0.0 & 0.1 & 0.6 & 0.1 & 0.0 \\
$\xfnufe$ &0.9 & 3.6 & 1.5 & 0.0 & 1.1 & 3.6 & 0.1 & 3.5 & 7.2 & 2.4 & 1.0 \\
$\xDbar$  &9.7 &21.7 & 1.9 & 0.2 &29.0 & 9.5 & 5.2 &14.4 &45.5 &10.9 & 4.1 \\
\end{tabular*}
\end{center}
\caption{
The relative errors $\De f/f$ (in percent) on the parton densities and
several structure functions, averaged over $x$.  The errors are
calculated at $\qs = 10$~\gev.  The columns give the individual
contributions from the sources described in \Se{se:errors}. The column
marked `total' gives the total error excluding the scale errors. 
}
\label{tab:errors}
\end{table}

\clearpage

%%-----------------------------------
\begin{figure} [p!]
\begin{center}
\includegraphics[bb = 20 120 560 690, scale = 0.8]{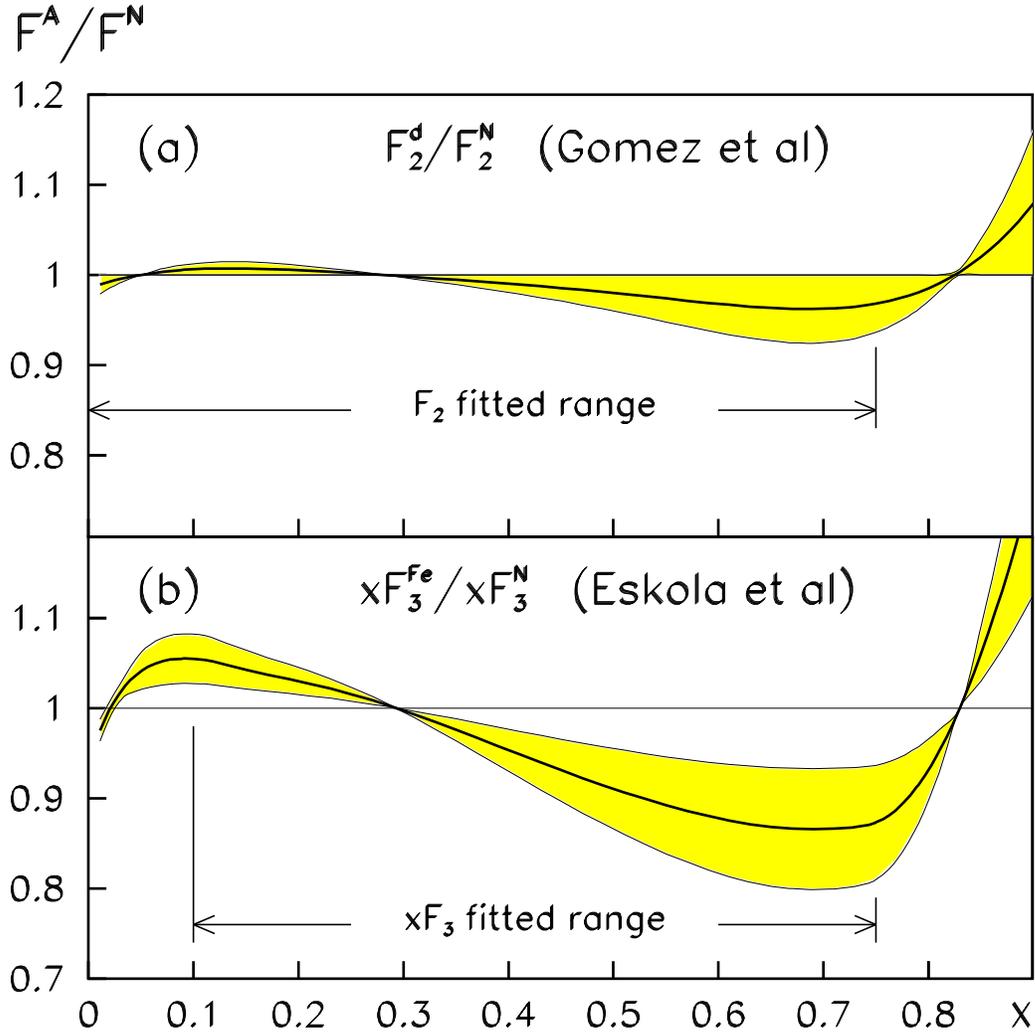}
\end{center}
\caption{
The ratio of nuclear structure functions to those of a free nucleon
versus $x$: (a) $\Ftd/F_2^N$ from~\cite{ref:gomez}; (b)
$\xfnufe/xF_3^{\nu N}$ from~\cite{ref:eskola}.  The shaded bands
correspond to the uncertainties on the ratios assumed in this analysis.
Also indicated are the $x$ ranges of the $F_2$ and $xF_3$ data included
in the QCD fit. 
}
\protect\label{fig:emc}
\end{figure}

%%-----------------------------------
\begin{figure} [p!]
\begin{center}
\includegraphics[bb = 60 150 500 700, scale = 0.95]{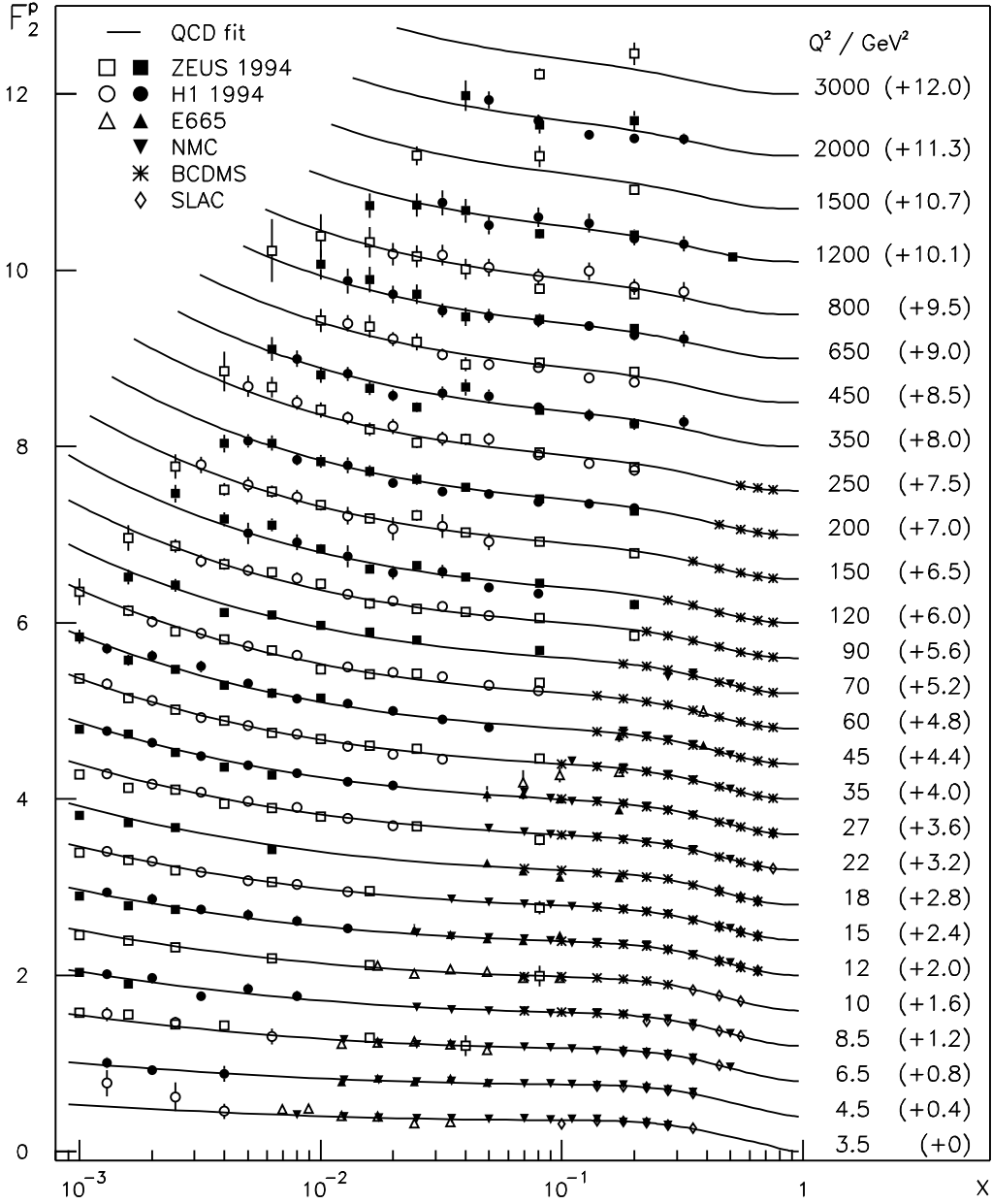}
\end{center}
\caption{
The proton $F_2$ structure function versus $x$ at fixed values of $\qs$
from ZEUS~\cite{ref:zeusnv}, H1~\cite{ref:h1qcd}, E665~\cite{ref:e665},
NMC~\cite{ref:newnmc}, BCDMS~\cite{ref:bcdms} and SLAC~\cite{ref:slac}.
Only shown are the data included in the QCD analysis. The full curves
show the result from the QCD fit.  For clarity the constants given in
brackets are added to $F_2$. The ZEUS, H1 and E665 data points are
plotted with open and solid symbols for alternating $\qs$ bins. 
}
\protect\label{fig:f2pvsx}
\end{figure}
%%-----------------------------------

%%-----------------------------------
\begin{figure} [p!]
\begin{center}
\includegraphics[bb = 20 160 580 670, scale = 0.81]{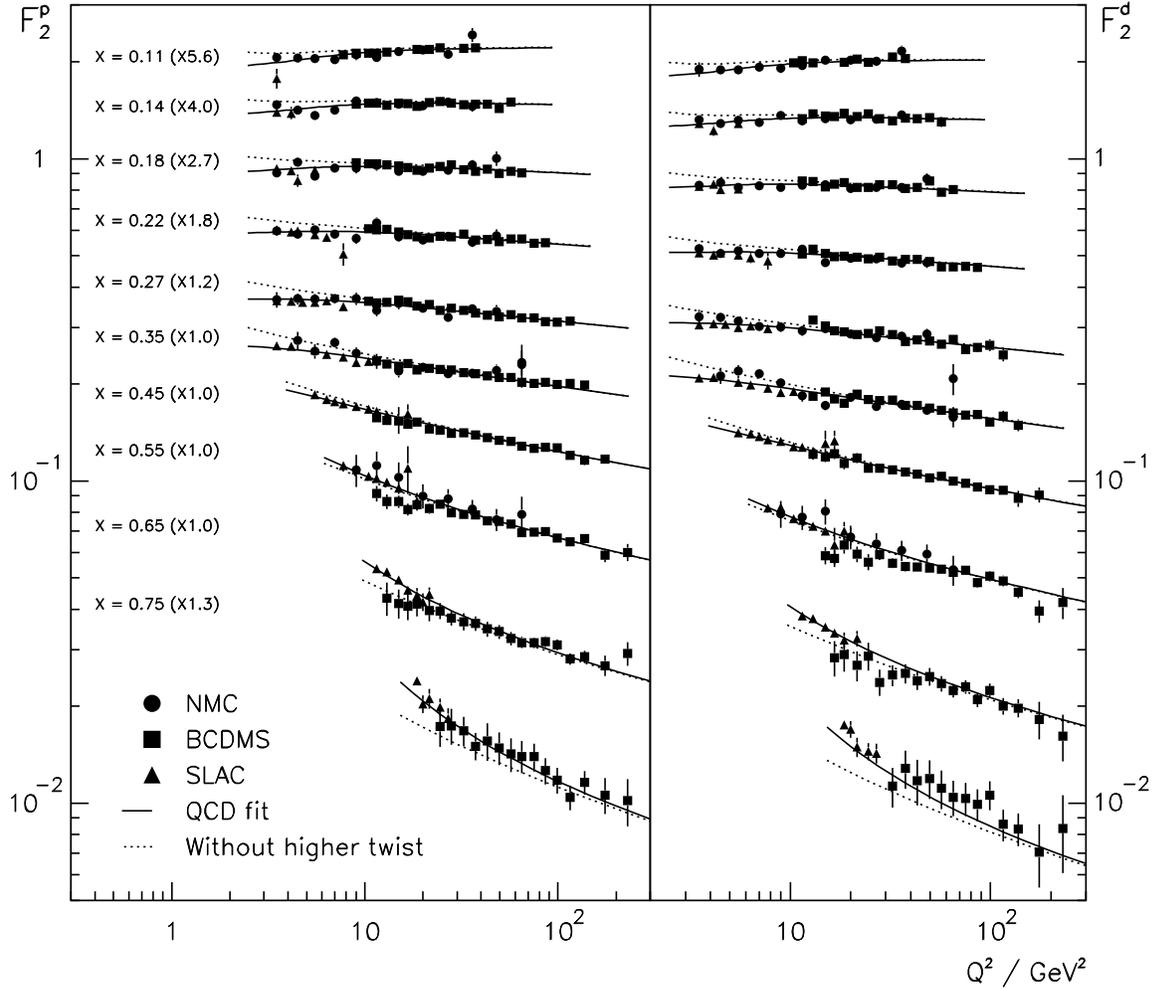}
\end{center}
\caption{
The structure functions $\Ftp$ (left) and $\Ftd$ (right) from fixed
target experiments~\cite{ref:newnmc,ref:bcdms,ref:slac} versus $\qs$ for
$x > 0.1$, $\qs > 3$ and $W^2 > 7$~\gev.  The full (dotted) curves
correspond to the QCD fit results including (excluding) higher twist
contributions. The $\Ftd$ curves include the correction for nuclear
effects in deuterium described in the text. The $x$ values and
multiplication factors (in brackets) given in the left-hand plot apply
to both $\Ftp$ and $\Ftd$. 
}
\protect\label{fig:f2vsq}
\end{figure}
%%-----------------------------------

%%-----------------------------------
\begin{figure} [p!]
\begin{center}
\includegraphics[bb = 50 140 520 700, scale = 0.95]{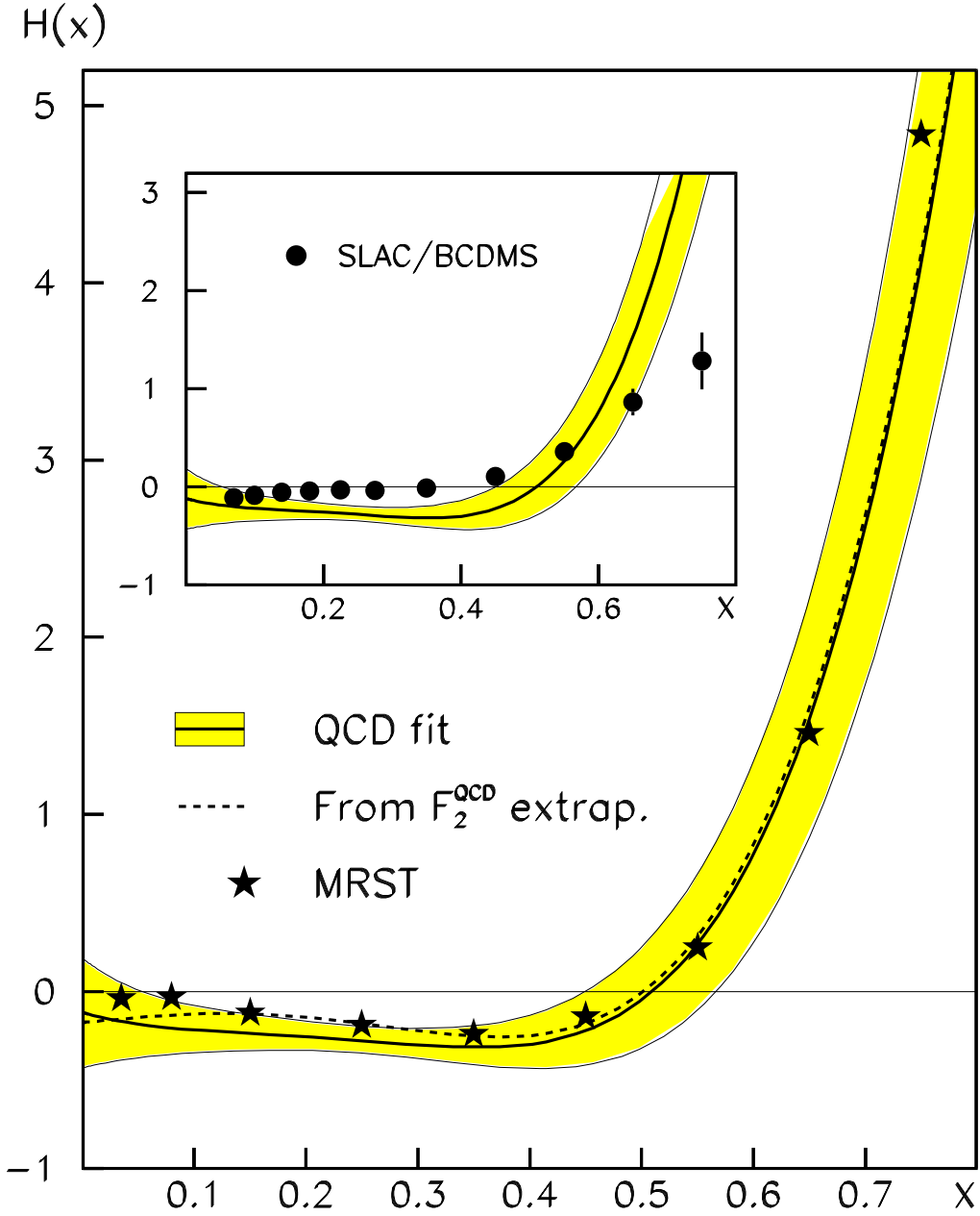}
\end{center}
\caption{
The higher twist correction $H(x)$ as defined in \Se{se:qcdfit}.  The
full curve corresponds to the result of the QCD analysis. The quadratic
sum of the error contributions (1) and (2) given \Se{se:errors} is drawn
as the shaded band around the curve. The dashed curve shows $H(x)$
obtained from an alternative fit described in \Se{se:checks}.  Also
plotted are the results from an analysis of MRST~\cite{ref:mrsht}
(asterisks).  In the inset $H(x)$ from this analysis is compared to the
result from the QCD fit of~\cite{ref:marcv} to the SLAC and BCDMS data. 
}
\protect\label{fig:htw}
\end{figure}
%%-----------------------------------

%%-----------------------------------
\begin{figure} [p!]
\begin{center}
\includegraphics[bb = 50 150 550 700, scale = 0.9]{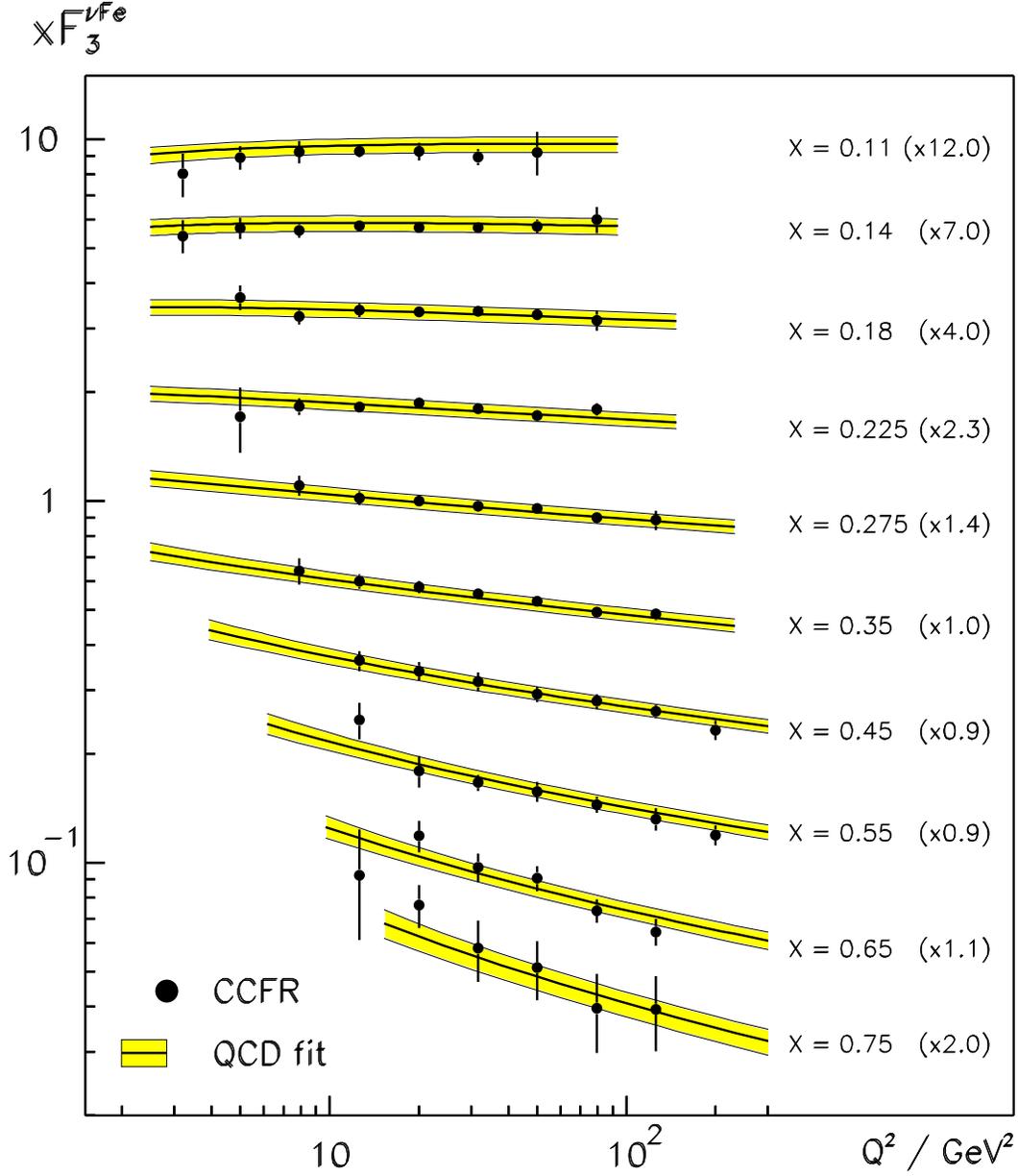}
\end{center}
\caption{
The $\xfnufe$ structure function from CCFR~\cite{ref:ccfrxf3} versus
$\qs$ at fixed values of $x > 0.1$.  The full curve shows the QCD
prediction corrected for nuclear effects on iron, described in the text.
The shaded bands indicate the error on the QCD fit.  For clarity $xF_3$
is multiplied by the factors indicated in brackets. 
}
\protect\label{fig:xf3vsq}
\end{figure}
%%-----------------------------------

\clearpage
 
%%-----------------------------------
\begin{figure} [p!]
\begin{center}
\includegraphics[bb = 40 180 520 700, scale = 0.9]{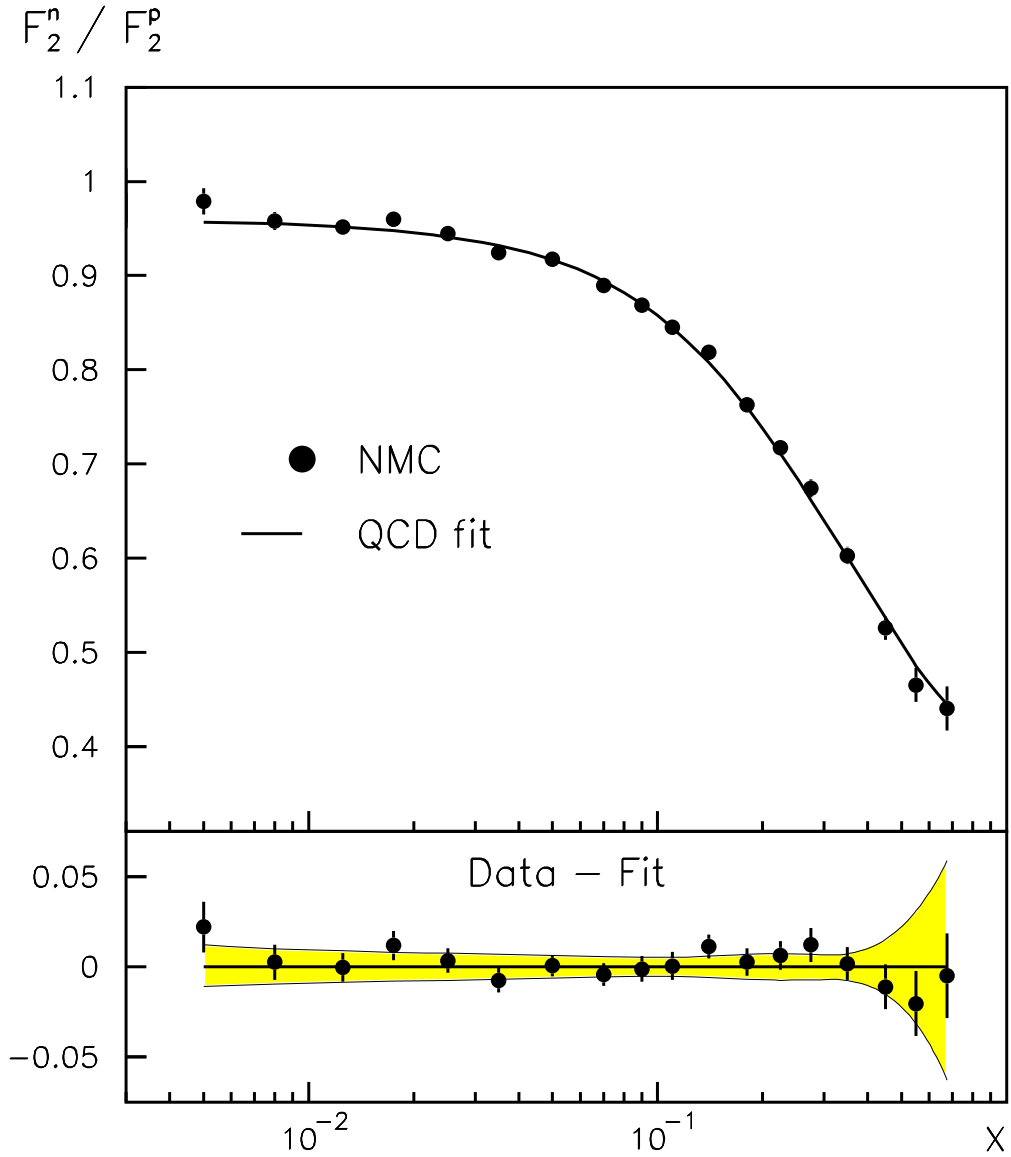}
\end{center}
\caption{
Top: The ratio $\Fnop$ versus $x$ from NMC, averaged over
$\qs$~\cite{ref:f2dop}.  The full curve shows the prediction from the
QCD fit, corrected for nuclear effects in deuterium as described in the
text. Bottom: The difference between the data and the QCD fit.  The
shaded band shows the error on the QCD prediction. 
}
\protect\label{fig:f2dop}
\end{figure}
%%-----------------------------------
 
%%-----------------------------------
\begin{figure} [p!]
\begin{center}
\includegraphics[bb = 30 140 520 720, scale = 0.9]{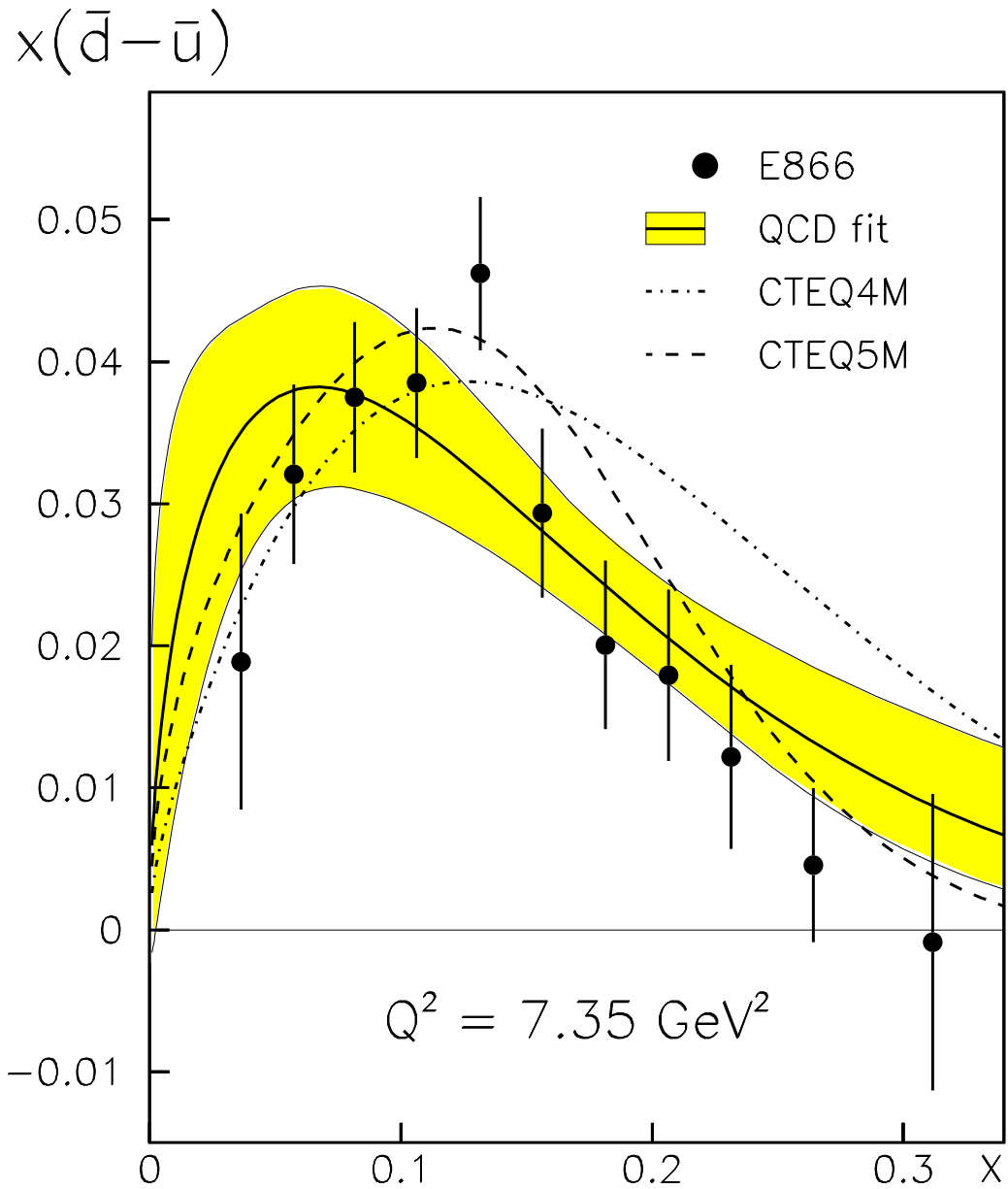}
\end{center}
\caption{
The difference $x(\bar{d}-\bar{u})$ from E866~\cite{ref:e866} versus $x$
at $\qs = 7.4$~\gev.  The full curve shows the result from the QCD fit
with the error drawn as the shaded band around the curve.  The dashed
(dashed-dotted) curve is the prediction from CTEQ4~\cite{ref:cteq4}
(CTEQ5~\cite{ref:cteq5}). 
}
\protect\label{fig:dmu}
\end{figure}
%%-----------------------------------

%%-----------------------------------
\begin{figure} [p!]
\begin{center}
\includegraphics[bb = 20 140 550 700, scale = 0.85]{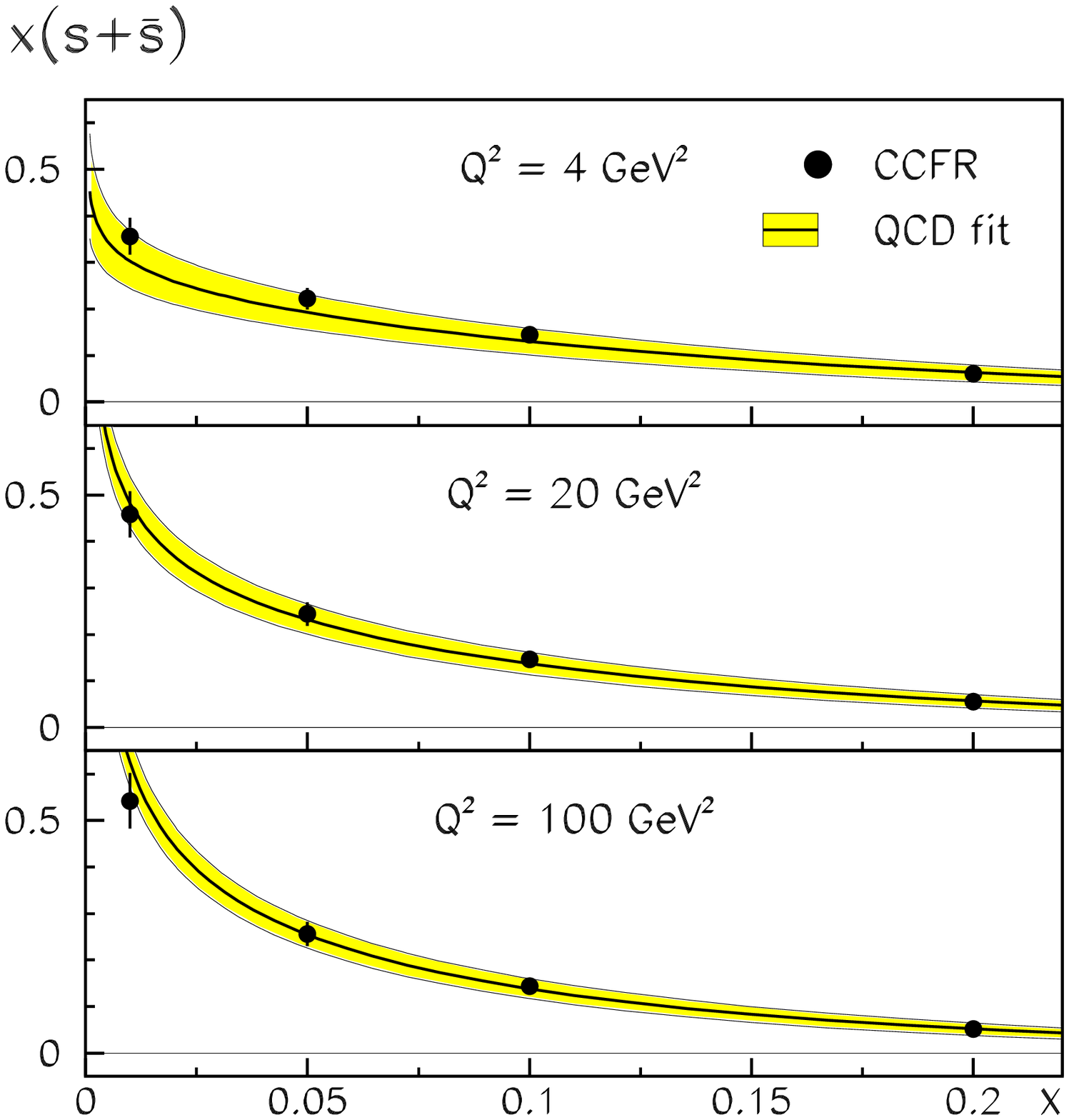}
\end{center}
\caption{
The strange quark density $x(s+\bar{s})$ versus $x$ at three 
values of $\qs$. The full circles show the data from
CCFR~\cite{ref:ccfrs}.  The curves correspond to the result from the QCD
fit. The error on the QCD prediction is indicated by the shaded bands. 
}
\protect\label{fig:strange}
\end{figure}
%%-----------------------------------
 
%%-----------------------------------
\begin{figure} [p!]
\begin{center}
\includegraphics[bb = 20 150 550 700, scale = 0.85]{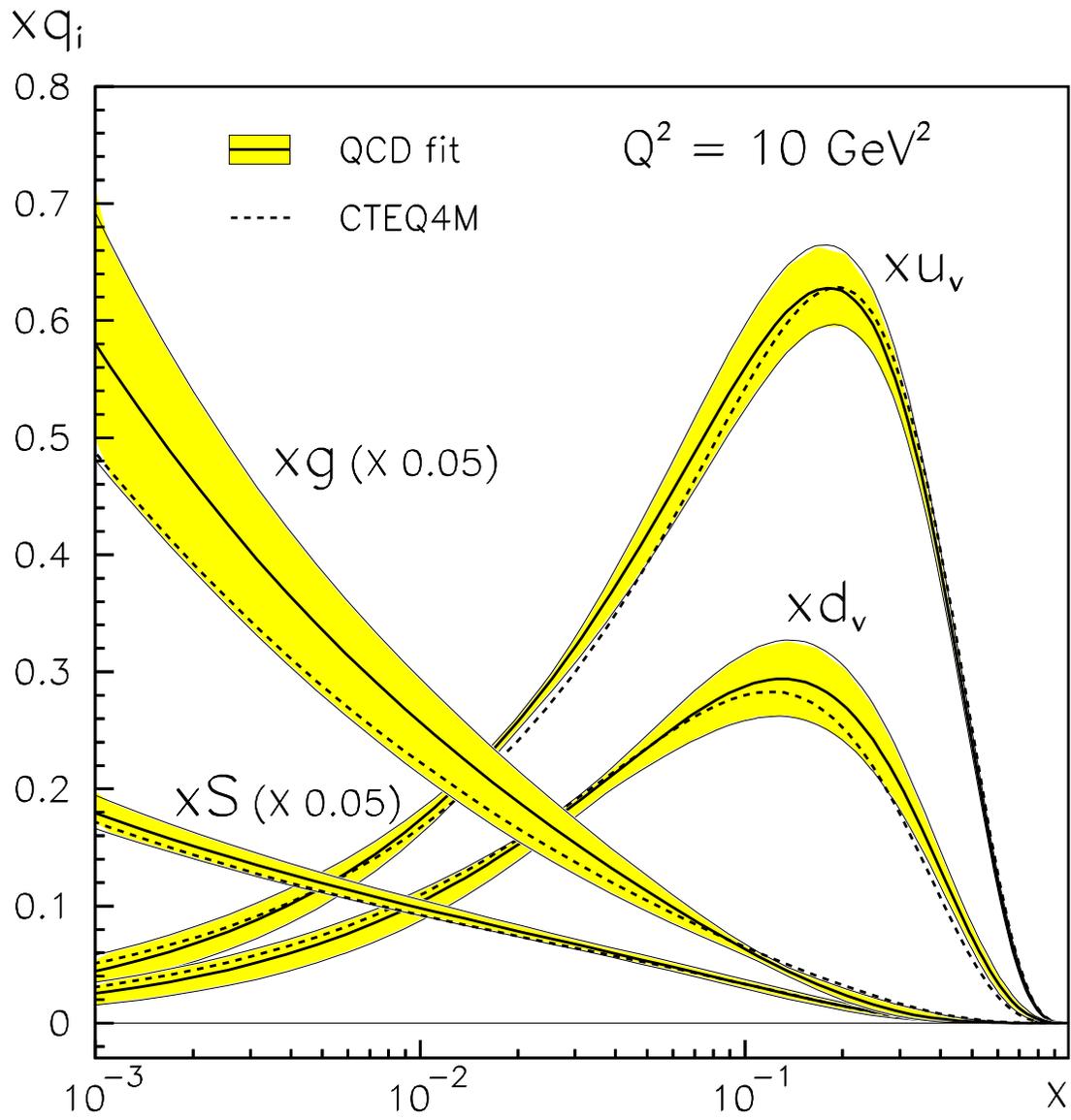}
\end{center}
\caption{
The parton momentum densities $xg$, $xS$ (both divided by a factor of
20), $xu_v$ and $xd_v$ versus $x$ at $\qs = 10$~\gev.
The full curves show the results from the QCD fit with the errors drawn
as shaded bands.  The dashed curves are from the CTEQ4~\cite{ref:cteq4}
parton distribution set. 
}
\protect\label{fig:pdf}
\end{figure}
%%-----------------------------------
 
%%-----------------------------------
\begin{figure} [p!]
\begin{center}
\includegraphics[bb = 35 140 540 710, scale = 0.85]{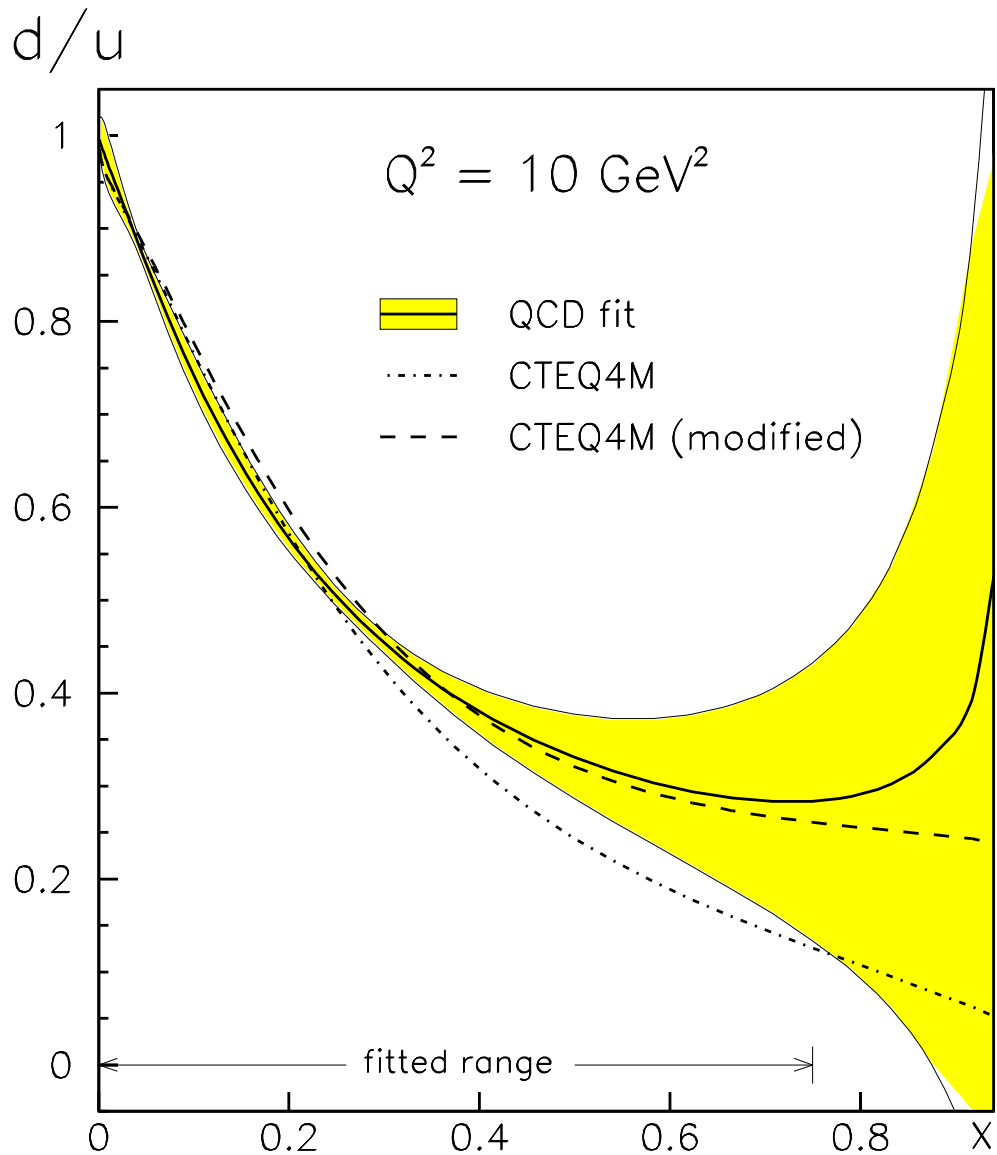}
\end{center}
\caption{
The ratio $(d+\bar{d})/(u+\bar{u})$ versus $x$ at $\qs = 10$~\gev\ from
the QCD fit (full curve) and CTEQ4~\cite{ref:cteq4} (dashed-dotted
curve).  The shaded band shows the error on the QCD fit. The dashed
curve corresponds to the CTEQ4 prediction with a modified down quark
density as described in the text. Indicated is the $x$ range covered by
the data included in the QCD fit. 
}
\protect\label{fig:dou}
\end{figure}
%%-----------------------------------

%%-----------------------------------
\begin{figure} [p!]
\begin{center}
\includegraphics[bb = 50 180 540 670, scale = 0.90]{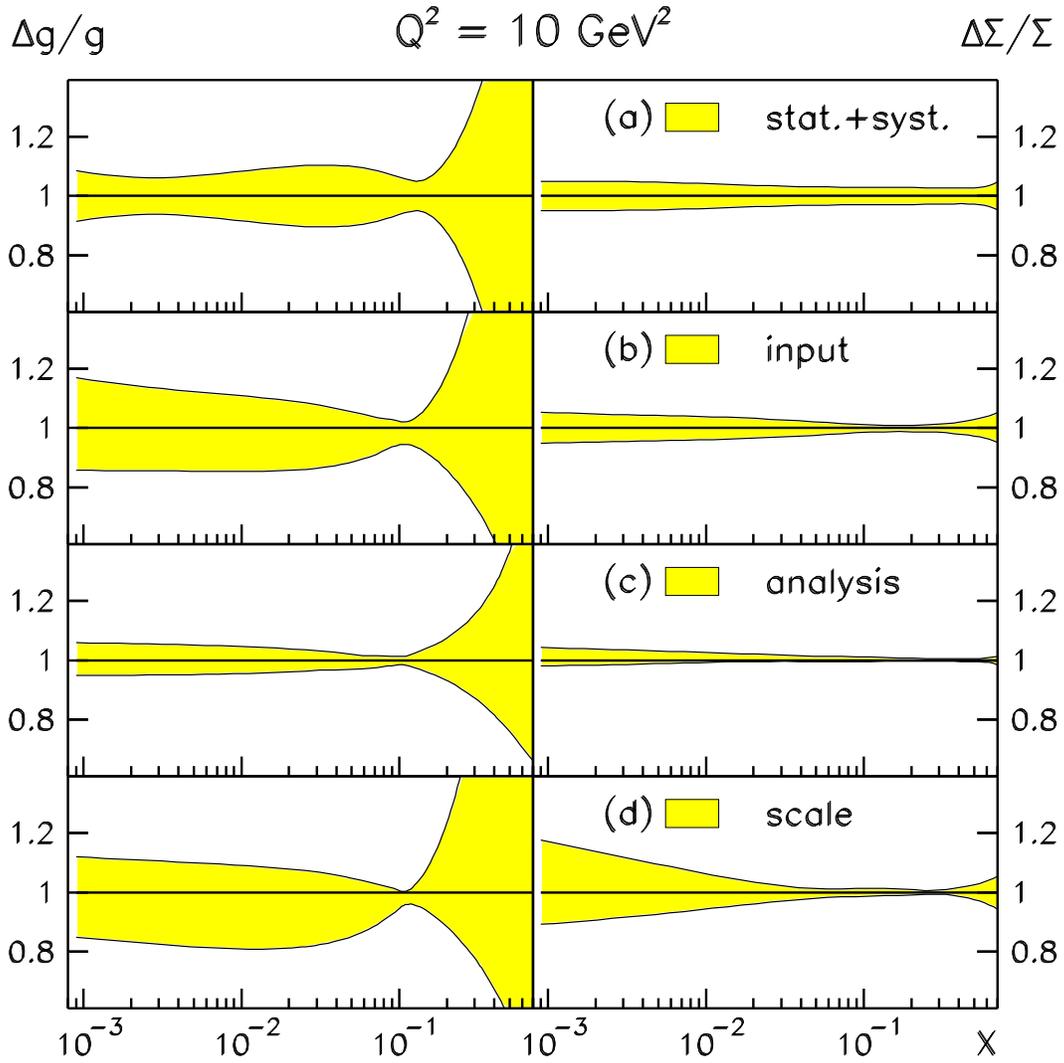}
\end{center}
\caption{
The four contributions to the relative error on the gluon density $\De
g/g$ (left hand plots) and the quark density $\De \Si / \Si$ (right hand
plots) versus $x$ at $\qs = 10$~\gev.  The error contributions are
defined in the text. 
}
\protect\label{fig:eband}
\end{figure}
%%-----------------------------------

%%-----------------------------------
\begin{figure} [p!]
\begin{center}
\includegraphics[bb = 30 150 550 690, scale = 0.85]{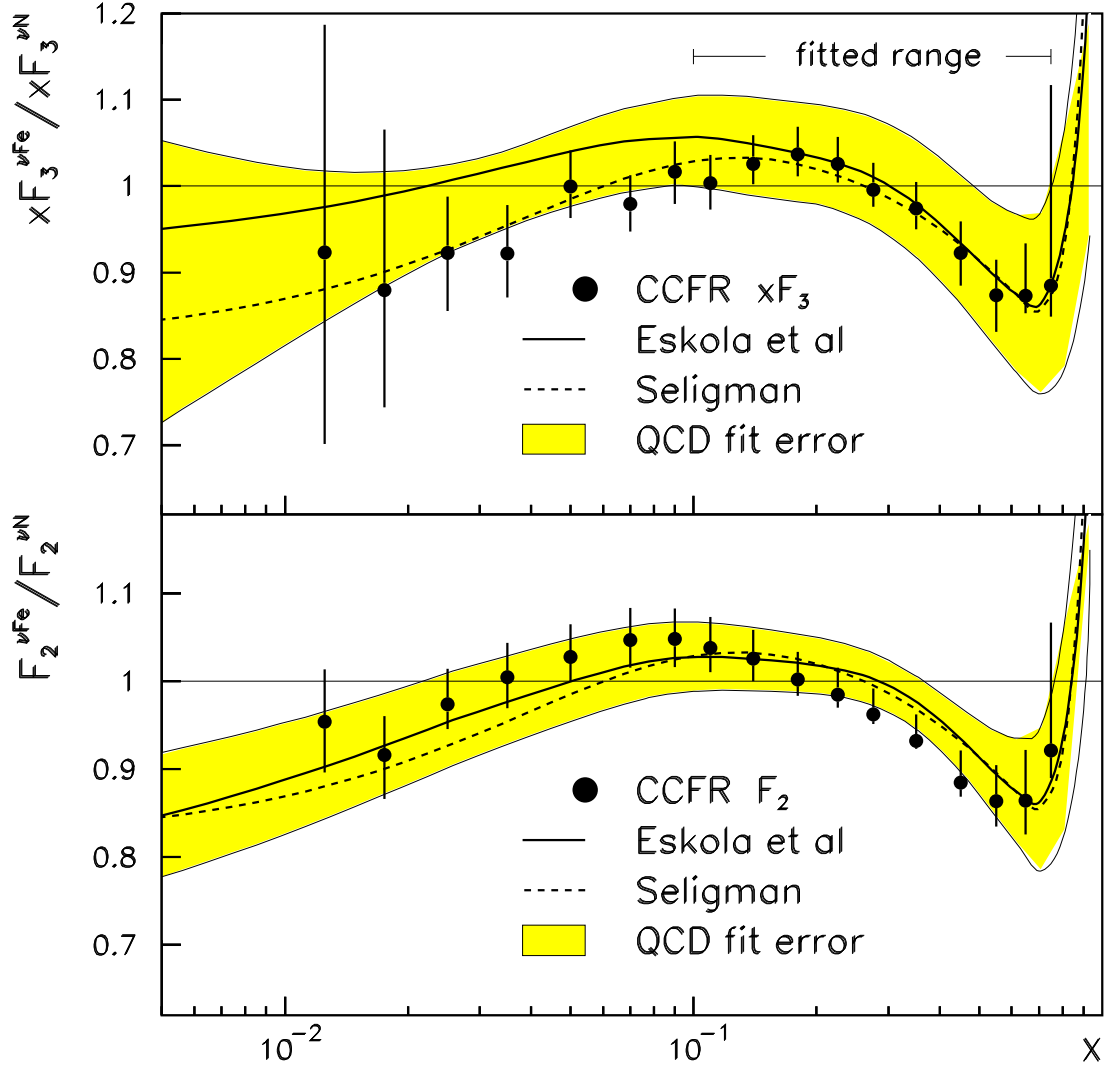}
\end{center}
\caption{
The ratios $\Rfthr$ (top) and $\Rftwo$ (bottom) of the CCFR $\nu$-Fe
data~\cite{ref:ccfrxf3} to the prediction from the QCD fit for neutrino
scattering on a free nucleon, obtained from a fit described in the text
(solid dots). The errors correspond to the statistical and systematic
errors on the data added in quadrature. Also shown are parameterizations
of nuclear effects in charged lepton scattering on iron from Eskola et
al.~\cite{ref:eskola} (full curves) and Seligman~\cite{ref:seligman}
(dashed curves). The error contributions from the QCD fit to $\Rfthr$
and $\Rftwo$ are drawn as the shaded bands around the full curves.  The
$x$ range of the $xF_3$ data included in the QCD analysis is indicated
in the upper plot. 
}
\protect\label{fig:emcrat}
\end{figure}
%%-----------------------------------

%%-----------------------------------
\begin{figure} [p!]
\begin{center}
\includegraphics[bb = 50 170 520 700, scale = 0.9]{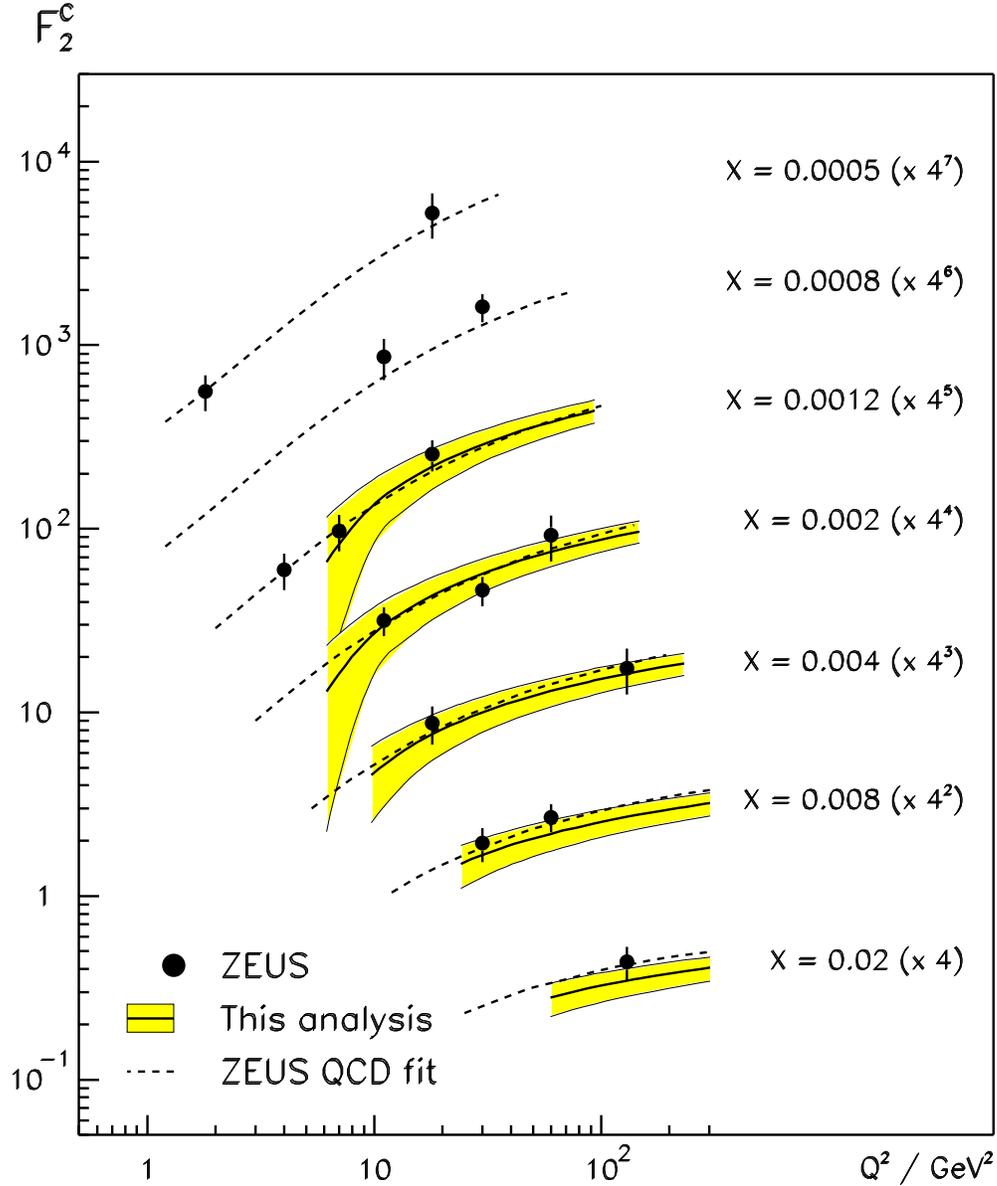}
\end{center}
\caption{ 
The charm structure function $F_2^c$ measured by ZEUS~\cite{ref:zeusfc}
versus $\qs$ at fixed values of $x \geq 5 \times 10^{-4}$.  The full
curves with shaded error bands show the predictions from the QCD fit.
These curves extrapolate to $F_2^c = 0$ at the charm threshold $\qs_c =
4$~\gev.  The dashed curves correspond to the predictions from the ZEUS
QCD fit~\cite{ref:zeusqcd} where charm mass effects were taken into
account.  For clarity $F_2^c$ is multiplied by the factors indicated in
brackets. 
}
\protect\label{fig:fcvsq}
\end{figure}
%%-----------------------------------

%%-----------------------------------
 
%%-----------------------------------
%%-----------------------------------
 
\end{document}